\documentclass[aps,pra,twocolumn,preprintnumbers,superscriptaddress,longbibliography,10pt]{revtex4-1}
\usepackage[T1]{fontenc} 
\usepackage[utf8]{inputenc}
\usepackage{multirow}
\usepackage{float}
\usepackage{bm}
\usepackage{booktabs} 
\usepackage{array} 
\usepackage{graphicx}
\usepackage{amsmath}
\usepackage{amssymb}
\usepackage{color}
\usepackage[per-mode=symbol,separate-uncertainty]{siunitx} 
\usepackage[capitalize]{cleveref} 
\usepackage{array,multirow,makecell}
\usepackage[table]{xcolor}
\usepackage{silence}
\usepackage{soul}
\usepackage{pifont}

\setcellgapes{1pt}
\makegapedcells
\newcolumntype{R}[1]{>{\raggedleft\arraybackslash }b{#1}}
\newcolumntype{L}[1]{>{\raggedright\arraybackslash }b{#1}}
\newcolumntype{C}[1]{>{\centering\arraybackslash }b{#1}}

\normalsize

\DeclareSIUnit{\belmilliwatt}{Bm}

\begin{document}

\author{Luca Planat}
\author{Arpit Ranadive}
\author{R\'emy Dassonneville}
\author{Javier Puertas Mart\'inez}
\author{S\'ebastien L\'eger}
\author{C\'ecile Naud}
\author{Olivier Buisson} 
\author{Wiebke Hasch-Guichard}
\affiliation{Univ. Grenoble Alpes, CNRS, Grenoble INP, Institut N\'eel, 38000 Grenoble, France}
\author{Denis M. Basko}
\affiliation{Laboratoire de Physique et Mod\'elisation des Milieux Condens\'es, Universit\'e Grenoble Alpes and CNRS, 25 rue des Martyrs, 38042 Grenoble, France}
\author{Nicolas Roch}
\affiliation{Univ. Grenoble Alpes, CNRS, Grenoble INP, Institut N\'eel, 38000 Grenoble, France}

\title{A photonic crystal Josephson traveling wave parametric amplifier}

\begin{abstract}
A microwave amplifier combining noise performances as close as possible to the quantum limit with large bandwidth and high saturation power is highly desirable for many solid state quantum technologies. Here we introduce a new Traveling Wave Parametric Amplifier based on Superconducting QUantum Interference Devices. It displays a \SI{3}{\giga\hertz} bandwidth, a \SI{-102}{\deci\belmilliwatt} 1-dB compression point and added noise near the quantum limit. Compared to previous state-of-the-art, it is an order of magnitude more compact, its characteristic impedance is \textit{in-situ} tunable and its fabrication process requires only two lithography steps. The key is the engineering of a gap in the dispersion relation of the transmission line. This is obtained using a periodic modulation of the SQUID size, similarly to what is done with photonic crystals. Moreover, we provide a new theoretical treatment to describe the non-trivial interplay between non-linearity and such periodicity. Our approach provides a path to co-integration with other quantum devices such as qubits given the low footprint and easy fabrication of our amplifier.
\end{abstract}

\maketitle

\section{Introduction}
A wide range of quantum technologies relies on the faithful detection of microwave signals at millikelvin temperatures. This includes quantum computing with superconducting~\cite{Vijay:2011gl} and spin~\cite{Stehlik:2015fs} qubits, electron spin resonance with quantum limited sensitivity~\cite{Bienfait:2015bo}, sensing the motion of nanomechanical resonators~\cite{Teufel:2011jg}, detecting cosmic microwave background using kinetic inductance detectors (KID)~\cite{Day:2003ck} or the quest for axionic dark matter~\cite{DuetalADMXCollaboration:2018jz}. For all these applications, sensitivity is limited by the first amplifier of the detection chain. When this level of performance is needed, Josephson parametric amplifiers (JPA)~\cite{Zimmer:1967fd,YURKE:1988vs} are considered as the technology of choice since they can reach the quantum limit of amplification~\cite{Caves:1982zz,Movshovich:1990bka,CastellanosBeltran:2008cg,Bergeal:2010iu}. These amplifiers are built around the parametric interaction between three waves called signal, pump and idler. However despite significant advances allowing improved bandwidth~\cite{Mutus:2013iw,Roy:2015ky} and saturation power~\cite{Eichler:2014iw,Liu:2017bi,Frattini:2018ud,planat2018understanding}, JPA's are still insufficient for demanding applications such as multiplexed readout of qubits~\cite{Heinsoo:2018db,Kundu:2019hb} or KID arrays~\cite{Monfardini:2010mv}. 

These limitations appear because the operating principle of a JPA is based on a resonant architecture. A solution to this is to engineer a non-resonant, non-linear device, such as a non-linear transmission line, where amplification arises step by step as was already extensively demonstrated in the optical domain~\cite{Agrawal:1616167} or with traveling wave tube amplifiers~\cite{TWTA}. Such devices are dubbed traveling wave parametric amplifiers (TWPA)~\cite{CULLEN:1958je}. They appear as the ultimate microwave amplifiers since they promise large gain, wide bandwidth, high saturation power and low noise approaching the quantum limit. However such appealing figures of merit come at a price. This non-linear transmission line should be impedance matched to \SI{50}{\ohm} to avoid unwanted reflections, synonymous of gain ripples. Moreover, not only should the energy between signal, pump and idler waves be conserved during the parametric process but also the momentum must be conserved as well since these are traveling waves. This latter condition, also known as phase matching~\cite{Agrawal:1616167}, is automatically fulfilled in a resonant device but is far less trivial to fulfill in a traveling wave structure. Indeed, the non-linearity at the heart of the amplification process also causes a non-linear phase-mismatch between the signal, idler and pump waves, which limits the gain and bandwidth of the amplifier. 

The first demonstration of a TWPA in the microwave domain relied on the non-linearity of NbTiN~\cite{Eom:2012kq}, a high kinetic inductance superconductor. The impedance matching was obtained using tapers while the phase matching was enforced via periodic loading structures, similarly to what is done with photonic crystals. However, such high kinetic inductance TWPA's (K-TWPA) show very small dispersion outside of the band of the periodic loading structures since their only cutoff frequency is the superconducting gap, located at frequencies as high as several hundreds gigahertz. This might give rise to shockwaves~\cite{5392514} and backward amplification~\cite{Erickson:2017dy}. Moreover, the non-linearity is intrinsically weak in such materials~\cite{Tholen:2007ea,Annunziata:2010,Maleeva:2018uya}. Also K-TWPA's require very high pump powers, which can cause unwanted heating effects~\cite{Eom:2012kq}. These two limitations probably explain why this device and subsequent implementations~\cite{Bockstiegel:2014bta,Vissers2016, Adamyan:2016bn, Chaudhuri:2017ij, Ranzani:2018fi} systematically showed noise close to but above the quantum limit. 

The first near quantum-limited TWPA~\cite{macklin2015near} was a non-linear transmission line made from a Josephson junction meta-material~\cite{Sweeny:1985km,YURKE:1996dt,Yaakobi2013parametric}. To achieve impedance matching, the high inductance of Josephson junctions was compensated by adding a discrete planar shunt capacitance to each unit cell. The phase matching issue was resolved using an elegant but technologically challenging resonant technique~\cite{Obrien2014resonant,white2015traveling}. This design still sets the state-of-the-art in term of gain, bandwidth and noise performance but requires a discouragingly complicated fabrication overhead such as multiple lithography steps and multiple metal layers. Moreover this amplifier cannot be directly integrated with superconducting qubits since it relies on a niobium tri-layer fabrication process~\cite{Tolpygo2015Fab}.

In this article, we introduce a new microwave Superconducting QUantum Interference Device (SQUID) based Traveling Wave Parametric Amplifier (STWPA). It combines near quantum-limited noise performances with simple design and fabrication process involving only two lithography steps~\cite{planat2019fab}, small foot-print, direct integration capability with superconducting qubits and in-situ flux tunability. Our device consists of SQUID meta-material~\cite{PuertasMartinez:2019gk}, which is subsequently covered by a thin dielectric and a top ground plane [see~\cref{fig1}({a})]. This simple approach allows us to solve the impedance matching issue. Moreover, contrary to previous TWPA's, this impedance can now be tuned in-situ since a  SQUID can be operated as tunable inductors controlled by a DC magnetic flux $\Phi$. To deal with the phase matching problem we engineer the dispersion relation of this non-linear transmission line using a spatial modulation of its impedance~\cite{Hutter2011josephson,Taguchi2015mode}. This periodic modulation opens a gap in the dispersion relation~\cite{1144771}, similar to what is observed for photonic crystals or for Bloch waves in crystal lattices [see~\cref{fig1}({c})]~\cite{Ashcroft}. In order to compensate for the non-linear phase matching intrinsic to the amplification process, the pump frequency is adjusted close to this gap to take advantage of the local distortion of the dispersion relation~\cite{Eom:2012kq}. Contrary to K-TWPA, our device benefits from a well controlled and stronger non-linearity~\cite{Weissl:2015,Krupko:2018is} and a cut-off or plasma frequency around \SI{30}{\giga\hertz}, which prevents unwanted non-linear phenomena and allows near quantum-limited performances. 

For quantitative understanding of the amplification process in our device we developed a novel self-consistent theory of the nonlinear pump propagation in the photonic crystal. When the pump wave propagates close to the edge of the photonic gap, it changes the properties of the photonic crystal (via the self-Kerr effect), modifies the spectral gap position and thus the pump intensity which enters the STWPA. Even for a weak spatial modulation and weak pump power, this effect is non-perturbative. The self-consistent description is crucial for quantitative modeling of the measured gain profile.

\section{Description of the device}
\label{sec:device}

\begin{figure}[htp]
\includegraphics[width=\linewidth]{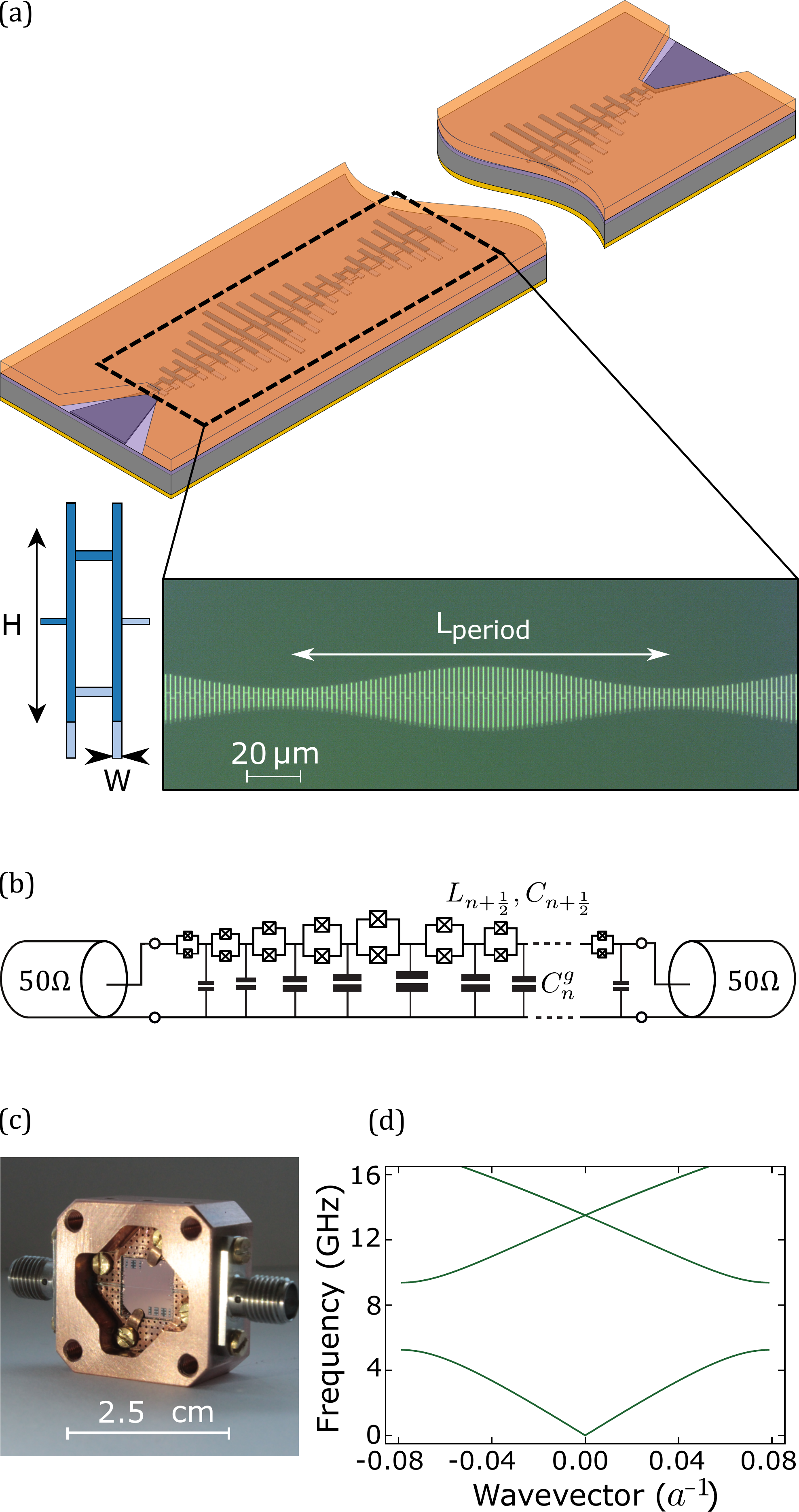}
\caption{(a)~Schematic of the photonic crystal STWPA. The spatially modulated SQUID array is fabricated on a silicon substrate. It is then covered by a thin alumina layer (purple, obtained via atomic layer deposition) and a thick copper layer orange, which is used as a top metallic ground. The inset shows the optical picture of two periods of the SQUID array. Only the height is modulated. The modulation amplitude on the figure ($\eta=60\%$) is greatly exaggerated for the sake of presentation; the actual samples have $\eta= 4\%$ and 2\%. (b)~Electrical sketch of the STWPA. The electrical parameters of the unit cells are following a sinusoidal modulation. (c)~Picture of the sample. The chip is the mauve square in the middle, plugged to a copper printed circuit board in its copper box. Magnetic field is applied normal to the chip using a coil directly anchored onto the copper box. (d)~Dispersion relation, plotted in the first Brillouin zone, of an array of 2160 SQUID with realistic microscopic parameters and a modulation amplitude $\eta=60\%$.}  
\label{fig1}
\end{figure}

In a first lithography step, we pattern an array of $N_\text{J}$ SQUIDs. We now want to match this array to \SI{50}{\ohm} characteristic impedance without adding extra components, such as large discrete capacitors. To do so we cover the array with a very thin dielectric layer (alumina, \SI{28}{\nano\meter}) and a thick copper layer to obtain a very close ground plane or equivalently to increase the ground capacitance per unit length. In a second lithography step we pattern openings in this ground plane to allow proper wire-bonding, as shown in~\cref{fig1}({a}). The obtained ground capacitance is a few tens of femtofarads per SQUID using this simple microstrip geometry~\cite{planat2019fab}. Finally, the absence of large capacitors for impedance matching and resonators for phase matching makes this STWPA very compact with a footprint of \SI{7.2}{\milli\meter} by \SI{13}{\micro\meter}. 

The two Josephson junctions forming each SQUID have a constant width $W$ and a periodically modulated height $H$, as schematically shown in Fig.~\ref{fig1} (see Appendix~\ref{app:fabrication} for details on the geometry). We chose sinusoidal modulations since closed form expressions regarding the amplitude and the position of the gap exist in this case~\cite{Hutter2011josephson}. Here we label the superconducting islands by an integer $n=0,1,\ldots,N_\text{J}$, so the SQUID connecting neighboring islands $n$ and $n+1$ is labeled by a half-integer $n+1/2$. We denote by $a$ the average length per SQUID, so that $N_\mathrm{J}a$ is the physical length of the array. 

\begin{table}
\begin{ruledtabular}
\begin{tabular}{|C{4.5cm}||C{1.5cm}|C{1.5cm}|}
\hline Device & A & B \\
\hline\hline 
Total SQUID number $N_\text{J}$ & 2160	 & 2184	\\
\hline 
Length per SQUID $a$ (\SI{}{\micro\meter}) & 3.3 & 3.3	\\
\hline 
Josephson capacitance $\bar{C}$ (\SI{}{\femto\farad}) & 485 & 485 \\
\hline 
Ground capacitance $\bar{C}^\mathrm{g}$ (\SI{}{\femto\farad}) & 42.6 & 42.6 \\
\hline 
Zero-flux inductance $\bar{L}$ (\SI{}{\pico\henry}) & $60.5\pm1.1$ & $61.7\pm1.1$ \\
\hline
Josephson modulation amplitude $\eta$ & 4\% 	& 2\% \\
\hline
Ground capacitance modulation amplitude $\zeta$ & 3.2\% 	& 1.6\% \\
\hline
Modulation period $N_\text{p}$ & 40	& 42	 \\
\hline 
Gap position 	(\SI{}{\giga\hertz})	& 7.45 & 7.15 \\
\hline 
Gap width (\SI{}{\mega\hertz})  & 350 & 200\\
\hline
\end{tabular}
\end{ruledtabular}
\caption{Parameters of samples A and B. The main difference is the modulation amplitude $\eta$, which reflects directly in the width of their photonic gap.}  
\label{tab1}
\end{table}

If $N_\text{p}$ is the number of SQUIDs per period, $G=2\pi/N_\text{p}$~is the reciprocal lattice vector (in units of~$a^{-1}$) corresponding to the modulation. Both the critical current of each SQUID $n+1/2$ (inversely proportional to its Josephson inductance $L_{n+1/2}$) and the SQUID capacitance $C_{n+1/2}$ are proportional to the junction area. The capacitance $C^\text{g}_n$ between each island~$n$ and the ground plane is proportional to the island area, and each island participates in two SQUIDS. 
Thus, the modulated circuit parameters can be written as
\begin{equation}\begin{split}
&L_{n+1/2}^{-1} = \bar{L}^{-1}\left[1+\eta\cos G(n+1/2)\right],\\
&C_{n+1/2} = \bar{C} \left[1+\eta\cos G(n+1/2)\right],\\
&C^\text{g}_n = \bar{C}^\text{g}\left[1+\frac\zeta{2}\sum_\pm\cos G(n\pm1/2)\right],
\label{eq:modulation}
\end{split}\end{equation}
where $\bar{L}$, $\bar{C}$, and $\bar{C}^\text{g}$ are the values corresponding to the average SQUID height while $\eta$ and $\zeta$ are modulation amplitudes of the Josephson parameters ($L$ and $C$) and the ground capacitance, respectively. The SQUID plasma frequency $1/\sqrt{L_{n+1/2}C_{n+1/2}}=1/\sqrt{\bar{L}\bar{C}}$ remains constant along the chain. This article presents two devices, labeled A and~B, whose parameters are listed in~\cref{tab1}. In our case, $\zeta$ is close to $\eta$. All parameters are determined from the geometry (see Appendix~\ref{app:fabrication} for details), except the Josephson inductance $\bar{L}$ which is found by fitting the experimental dispersion relation in the linear region, as discussed in the next section.

To model the device theoretically, we use the classical equations of motion for the superconducting phases $\phi_n$ on each island~$n$,
\begin{align}
&\frac{d^2}{dt^2}\left[C^\text{g}_{n}\phi_n+
C_{n+1/2}\left(\phi_n-\phi_{n+1}\right)
+C_{n-1/2}\left(\phi_n-\phi_{n-1}\right)\right]
\nonumber\\
&{}+\frac{\sin(\phi_n-\phi_{n+1})}{L_{n+1/2}}
+\frac{\sin(\phi_n-\phi_{n+1})}{L_{n-1/2}}=0.
\label{eq:maineq=}
\end{align}
The linear dispersion is obtained by plugging into these equations the modulated parameters as given by Eq.~(\ref{eq:modulation}), expanding $\sin\theta\approx\theta$, and looking for the solution in the form
\begin{equation}\label{eq:solution}
\phi_n(t)=\left(Ae^{ikn}+Be^{i(k-G)n}\right)e^{-i\omega{t}}+\mbox{c. c.}.
\end{equation}
The nonlinear pump propagation and the gap shift are determined by expanding $\sin\theta\approx\theta-\theta^3/6$ in Eqs.~(\ref{eq:maineq=}), substituting solution~(\ref{eq:solution}), retaining only terms proportional to $e^{\pm{i}(kn-\omega{t})},e^{\pm{i}(kn-Gn-\omega{t})}$, and solving self-consistently for $A,B,k$ for a given incident pump power (see Appendix~\ref{app:theory} for details). The signal amplification is obtained by adding weak signal and idler waves $\phi_n^\text{s}e^{-i\omega_\text{s}t}$, $\phi_n^\text{i}e^{-i\omega_\text{i}t}$ on top of the previously determined pump solution, linearizing Eqs.~(\ref{eq:maineq=}) with respect to $\phi_n^\text{s,i}$, and solving the resulting linear equations directly, without any \textit{a priori} assumptions on the spatial dependence of  $\phi_n^\text{s,i}$.

\section{Transmission in a nonlinear photonic crystal}
\label{sec:transmission}

\begin{figure}[htp]
\includegraphics[width=\linewidth]{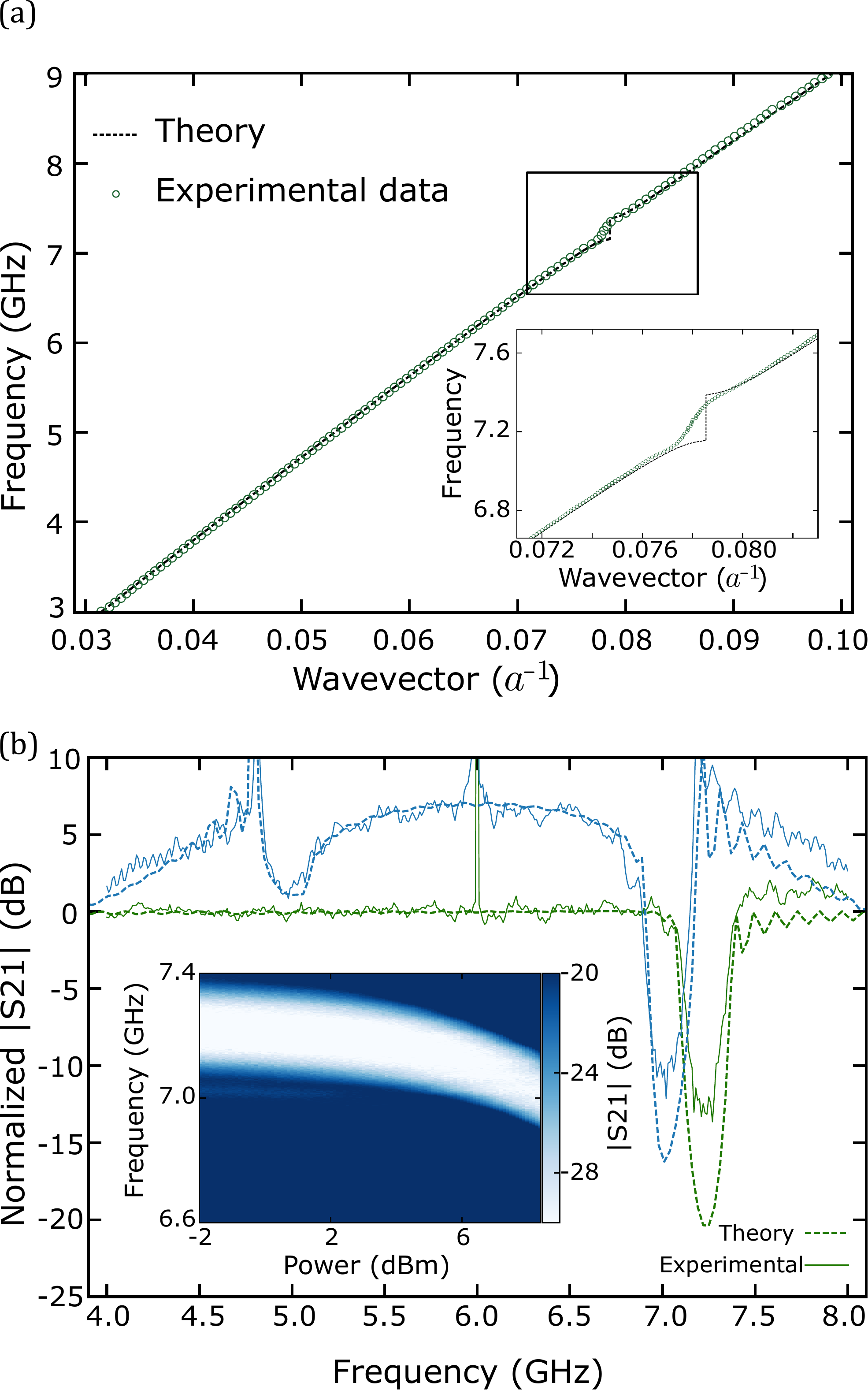}
\caption{(a)~Dispersion relation of sample A measured at very low magnetic flux ($\Phi/\Phi_0=0.1$). Theoretical curve is a fit to the linear dispersion relation with $\bar{L}$ as a fit parameter. The kink at \SI{7}{\giga\hertz} is the photonic gap created by spatial modulation. \textbf{Inset}. Close-up of the photonic gap. Theory agrees with experimental data and reproduces gap position with a 0.6\% error. (b)~Normalized probe transmission of sample A for different pump powers ($\omega_\text{P}/2\pi= \SI{6}{\giga\hertz}$, {green: $P_\text{P}=\SI{-83.8}{\deci\belmilliwatt}$, blue: $P_\text{P}=\SI{-70.3}{\deci\belmilliwatt}$}). The gap is shifted by more than its width at large pump power. \textbf{Inset}. Color map of the probe transmission versus probe frequency and pump power.}
\label{fig2}
\end{figure}

We first measure the dispersion relation in the linear regime without the pump tone and at very low signal power. The wavevector $k$ is obtained by measuring the phase along the STWPA, dividing by its full length $L_\text{twpa}=N_\text{J}a$, and subtracting the phase measured with a dummy \SI{50}{\ohm} line from it (See \cref{app:setup}). The $2\pi$ uncertainty introduced by the absence of signal for very low frequencies is lifted by following the procedure described in~\cite{macklin2015near}. Namely, we fitted the linear region at low frequencies with an affine law in order to get the intercept, which is an integer multiple of $2\pi$. We then substract this intercept. Once this calibration done, we can fit the dispersion relation with our model, leaving $\bar{L}$ as the only fit parameter, while the rest of the parameters ($\eta$, $N_\text{J}$, $N_\text{p}$, $\bar{C}$, $\bar{C}^\text{g}$) are set by the design and the fabrication process. We observe a quasi linear dependence within the $3-\SI{9}{\giga\hertz}$ band and a kink around \SI{7}{\giga\hertz} corresponding to the photonic gap [see \cref{fig2}({a})]. Inset of~\cref{fig2}~({a}) shows a zoom-in on the gap region of sample A. The gap is located near \SI{7.25}{\giga\hertz} and is about \SI{350}{\mega\hertz} wide. In the rest of the article, all microscopic and fabrication parameters are fixed.

Although there is an overall good agreement between theory and experimental data, there is a little discrepancy regarding the exact wavevector and frequency position of the gap. According to the theory, its wavevector position should be $G/2=7.85\times10^{-2}$ and its frequency \SI{7.29}{\giga\hertz}. Experimentally, we found $7.80\times10^{-2}$ and \SI{7.24}{\giga\hertz}, which is a deviation of 0.6\% and 0.7\% respectively, from the expected value. This should be understood as the precision of our experimental way to obtain the dispersion relation. To obtain a perfect frequency match of the gap position, we allow up to 1.75\% $\bar{L}$ variation.  

An important feature of this photonic crystal STWPA is the dependence of the gap position on the incident pump power. Understanding this non-linear shift is of prime importance since the pump frequency needs to be precisely adjusted close to the gap whose frequency is shifted by the pump power at the same time. There are two ways to investigate this effect. The first one, that we call \textit{self-shift}, consists in sending an intense microwave tone around the gap frequency and measuring transmission. Alternatively, we can infer the \textit{cross-shift} by pumping the STWPA far from the gap and by measuring its transmission using a probe tone. Experimentally it is more difficult to access the \textit{self-shift} since the measurement time has to be matched to the intrinsic dynamics of this non-perturbative phenomena, in a way similar to what is done for JPA close to bifurcation~\cite{Vijay:2009wp}. Hence, we focused on cross-shift measurements.
 
\begin{figure*}[htb]
\includegraphics[width=\textwidth]{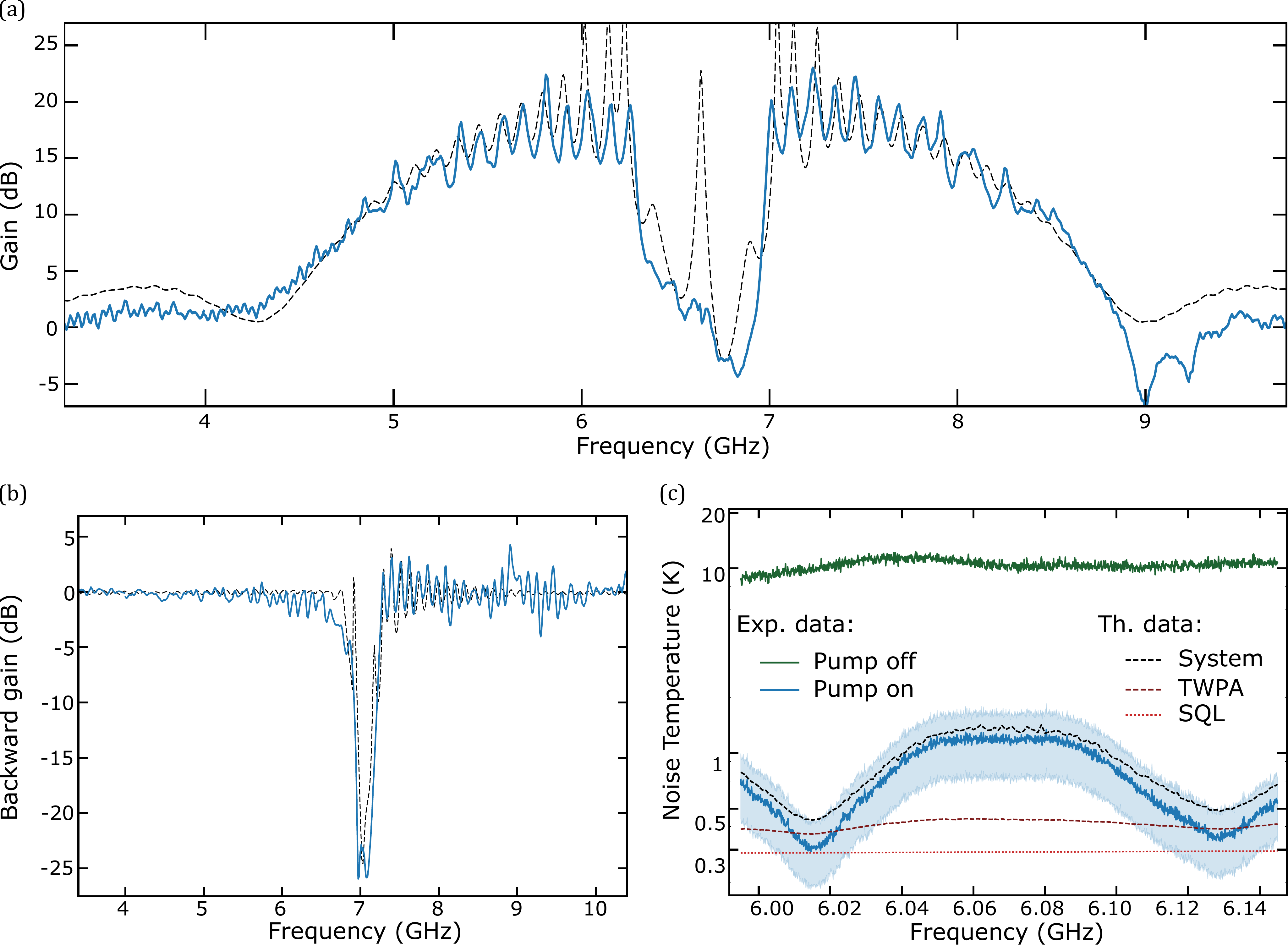}
\caption{(a)~Gain profile of the STWPA at magnetic flux $\Phi/\Phi_0=0.2$ and pump frequency \SI{6.635}{\giga\hertz}. Theoretical gain profile (dashed black line) is obtained for a pump power $P_\text{input}=\SI{-70.2}{\deci\belmilliwatt}$. (b)~Backward gain measured with $\Phi/\Phi_0=0.1$ and pump frequency \SI{6.9108}{\giga\hertz}. Expected gain is plotted for a pump power  $P_\text{input}=\SI{-71.4}{\deci\belmilliwatt}$. (c)~Noise temperature of the STWPA. Measured PSD normalised by the system gain when the STWPA pump is off (green solid line) or on (blue solid line, the light blue area corresponds to error bars). The red dotted line corresponds the standard quantum limit (SQL) of noise. The brown dashed line refers to the intrinsic noise of the STWPA, which is slightly above the SQL because of internal losses. The black dashed line is the total system noise temperature. It is higher than the STWPA intrinsic noise temperature because of the finite STWPA gain. Noise ripples are caused by STWPA gain ripples.}
\label{fig3}
\end{figure*}

Probe transmission while STWPA is pumped at \SI{6}{\giga\hertz} is shown on~\cref{fig2}({b}). At low pump power we recover the linear transmission. At high power, we observe up to \SI{7}{\decibel} gain and a \SI{-3}{\decibel} bandwidth of \SI{1.5}{\giga\hertz}. These small gain and bandwidth are the consequences of poor phase-matching since the pump does not benefit of the dispersive feature provided by the gap. More interestingly, we observe a large shift of the gap frequency position compared to the low pump power case. This nonlinear shift has the same order of magnitude as the actual gap width (few hundreds of megahertz).

Comparison between data and theory is done by setting all the parameters of the meta-material to the values obtained by the dispersion relation analysis. The only free parameter is the pump power. Since this model reproduces successfully the nonlinear behavior of our STWPA over a broad pump power range, it provides an accurate method to calibrate the attenuation of the input line. The calibration of the input attenuation line has been carried out by taking advantage of the large cross-shift of the gap frequency position with respect to the pump power. We compared the experimental pump power sent from the RF source at room temperature to the theoretical pump power at the input of the STWPA needed to make the photonic gap shift from its initial frequency. The difference between the experimental 'hot' pump power and the theoretical 'cold' pump power gives an attenuation of \SI{78.8}{\decibel} between the RF source output and the STWPA input. This value is the sum of the input line attenuation $A_\text{system}$ and the attenuation of the pump amplitude within the STWPA $A_\text{twpa}$. Given the small loss of the STWPA, it is taken as position independent~\cite{macklin2015near}, such that $A_\text{twpa}=\exp(k_\text{p}''N_\text{J}a/2)$, where $k_\text{p}''=k_\text{p}'\tan(\delta/2)$ is the imaginary part of the complex wavevector (in the limit of small losses), $N_\text{J}a$ is the total length of the STWPA and $\tan{\delta}$ is the loss tangent. We find $A_\text{twpa}=\SI{1.8}{\decibel}$ at \SI{6}{\giga\hertz}. This result, giving $A_\text{system}=\SI{77.0}{\decibel}$, is in good agreement with the system attenuation calculated at \SI{6}{\giga\hertz} by summing all the discrete attenuators (\SI{63.0}{\decibel}), power splitter (\SI{6.8}{\decibel}), filter (\SI{2.0}{\decibel}), cryogenic temperature cables (\SI{80}{\centi\meter}, \SI{1.5}{\decibel}) and room temperature cables (\SI{2.3}{\decibel}), giving \SI{75.6}{\decibel} and leading to a discrepancy of \SI{1.4}{\decibel}, which can be seen as the uncertainty of our calibration method. We then decided to set the system attenuation to $A_\text{system}=\SI{77.0}{\decibel} \pm \SI{1.4}{\decibel}$. Attenuation calibration is described in details in Appendix~\ref{app:input}.

\section{Phase matched amplification}
We now report on the characteristics of the STWPA when pumped close to the edge of the photonic bandgap (Fig.~\ref{fig3}). Larger gain and bandwidth are expected since the non-linear phase shift is now compensated. Finding the correct pump power and frequency is slightly trickier than for resonant phase matching (RPM) JTWPA~\cite{macklin2015near,white2015traveling} since the gap moves when pump power is changed. However, this issue is easy to overcome: similarly to current pumped JPA, the optimal pump configuration (frequency and power) must be found incrementally. Whenever power is increased, the frequency must be lowered to compensate its own back-action until optimal gain is reached. 

With sample A, we obtain a mean maximum gain around \SI{18}{\decibel} over a \SI{3}{\giga\hertz} bandwidth, from which \SI{700}{\mega\hertz} must be taken out given the photonic gap. We also observe up to \SI{5}{\decibel} gain ripples. They mainly come from imperfect impedance matching between the STWPA and the measurement setup. Theoretical and experimentally achieved gain are shown in~\cref{fig3}({a}). Here, attenuation was again the only free parameter (\SI{79.5}{\decibel} at \SI{6.6}{\giga\hertz}) and the measured gain is in very good agreement with theory. Sample B shows smaller gain and bandwidth [see~\cref{fig4}({b})] because of its smaller gap. This feature is again well captured by our theoretical formalism. A qualitative explanation is that a small curvature around the gap can only provide a limited correction to the nonlinear phase-shift induced by the strong pump~\cite{white2015traveling}. In the present case, curvature and size of the gap are directly linked since we used a simple sinusoidal modulation. However, in principle, a more advanced gap engineering could allow larger curvature without scarifying to the gap size.

We further investigated the behavior of our STWPA by probing backward amplification, that is amplification when signal and pump propagate in opposite directions.  Backward amplification could potentially happen in periodic structure such as our STWPA and could be detrimental to the behavior of the device, creating gain ripples and reducing threshold for parametric oscillations~\cite{Obrien2014resonant,Erickson:2017dy}. But more importantly this could affect the STWPA directionality and thus create a strong back-action on the measured devices. 

Choosing a conservative approach, we optimized the pumping parameters to maximize backward amplification. 
Probing backward gain was carried out with configuration \ding{193}, discussed in~\cref{app:setup}. This configuration allows to send the STWPA pump in one direction while measuring the signal propagating in the opposite direction. We first probed the signal gain in the standard way (pump and signal have the same propagation direction) to get an optimal set of pump frequency $\Omega_\text{opti}$ and power $P_\text{opti}$, giving $\bar{G}_\text{opti}=\SI{18}{\decibel}$. Pump tone is then sent through the STWPA output while keeping the pump frequency fixed to $\Omega_\text{opti}$. Signal gain is then probed while sweeping pump power, since attenuation from one line to another can change. 

The highest backward gain we could obtain is shown in~\cref{fig3}({b}) and stays below \SI{5}{\decibel}. We attribute this small value to two features of our STWPA. First, we achieved a good impedance matching between the SQUID array and the input and output lines. Second, the dispersion relation is engineered with a low plasma frequency of $\SI{30}{\giga\hertz}$. Then the parasitic parametric processes coming from higher Bloch bands are strongly suppressed since they are poorly phase matched. As a last check, we compared the experimental data to the theoretical backward amplification [see~\cref{fig3}({b})]. Again, theory and experimental data are in very good agreement.

\section{Noise measurements}
Noise performance of the STWPA are estimated by measuring the noise power spectral density (PSD) at the output of the whole measurement chain, when the pump is on and when it is off. We relate the measured noise PSD to the system noise $T_\text{N,syst}$ as:
\begin{equation}
\text{PSD} = G_\text{syst}k_\text{B}T_\text{N,syst} 
\end{equation}     
where $G_\text{syst}$ is the system gain. We obtain $G_\text{syst}$ using an accurate calibration of the input line (see Sec.~\ref{sec:transmission}). The system noise temperature is shown in~\cref{fig3}({c}). It has two main contributions, finite STWPA gain and intrinsic losses within the STWPA. The first one causes oscillations in $T_\text{N,syst}$. Indeed, when the STWPA gain decreases, the noise contribution of the following High Electron Mobility Transistor (HEMT) amplifier increases. The second one directly affects the STWPA noise temperature since signal is lost while traveling inside the device. We measured a system noise temperature very close to the standard quantum limit and in agreement with the modeling of our device (see~\cref{app:noise}).

STWPA intrinsic input noise is plotted in~\cref{fig3}({c}) (brown dotted line, in kelvin). The system noise [black dotted line,~\cref{fig3}({c})] is calculated as:
\begin{equation}\label{eq:Nsystem}
N_\text{system}=N_\text{twpa} + N_\text{system,off}/G_\text{twpa}
\end{equation}
where $N_\text{system,off}$ is the measured system noise when the STWPA pump is off [green dotted line, \cref{fig3}({c})].
The good agreement between the measurement and the modeling shows that our STWPA intrinsic noise [\cref{fig3}({c}), brown dashed line] is just above the standard quantum limit (red dashed line).
 
We characterized as well the signal-to-noise ratio (SNR) improvement and gain compression thanks to the line calibration mentioned earlier. SNR improvement is carried out by sending a very low power signal into the system while probing the power spectral density with a spectral analyser with a bandwidth of \SI{20}{\kilo\hertz} when the STWPA pump is on and off. Data are normalized with the total system gain. SNR improvement has been taken while signal is sent on a gain peak and a gain dip. 
We observed an improvement between \SI{15}{\deci\bel} (gain peak) and \SI{10}{\deci\bel} (gain dip) in the SNR for an input signal $P_\text{signal}=\SI{-135}{\deci\belmilliwatt}$ while we measured a 1dB compression point $P_\text{1dB}=\SI{-102}{\deci\belmilliwatt}$ (see Appendix~\ref{app:SNR}).  

\section{Response to a flux of the SQUID transmission line}
\label{sec:flux}

\begin{figure}
\includegraphics[width=\linewidth]{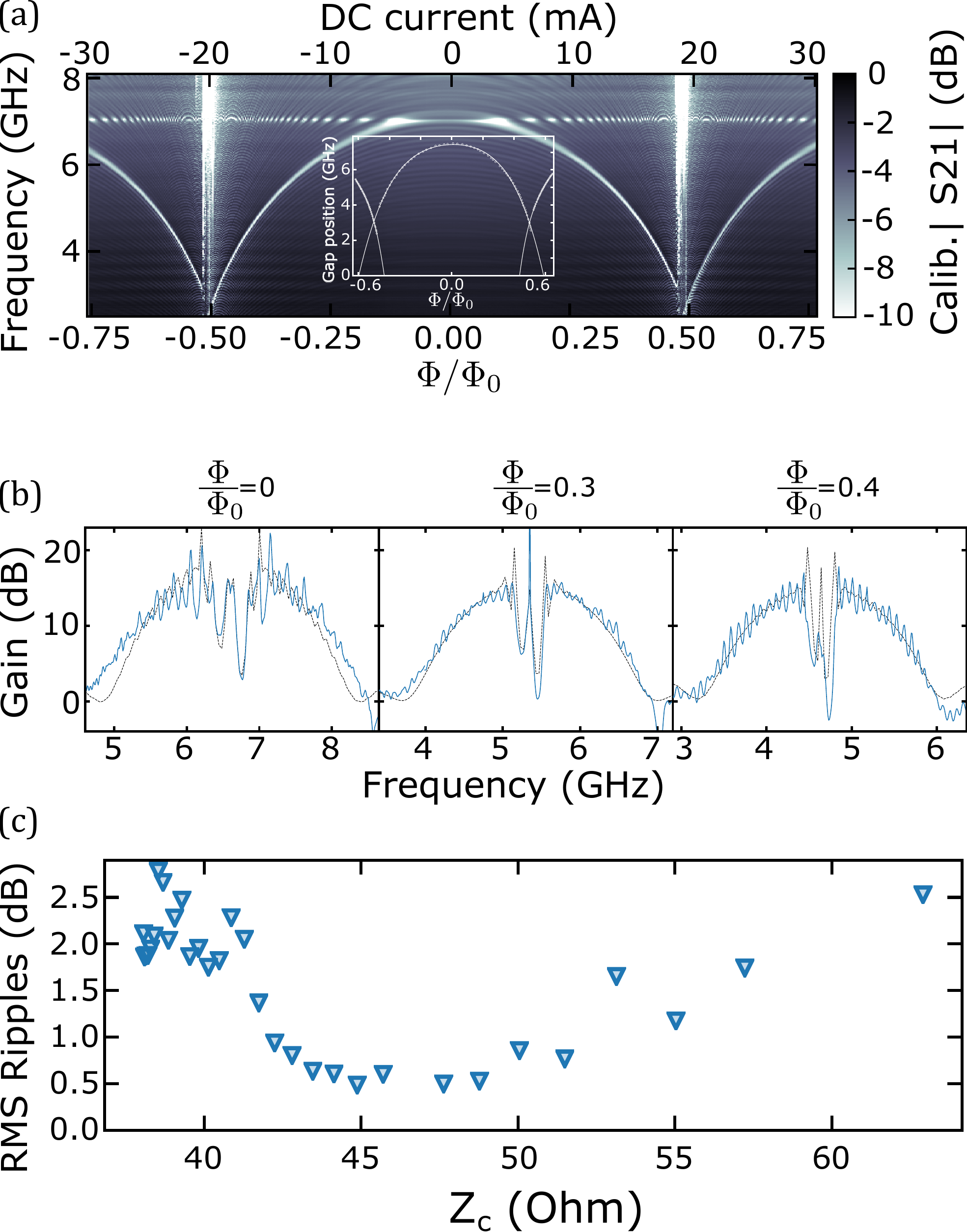}
\caption{(a) Calibrated transmission versus flux and probe frequency. Inset: interpolation of the gap position down to zero frequency. The distance between two inter-crossing domains at zero frequency gives information about the SQUID stray inductance. (b)~Three gain profiles of device B for three different flux points. Gain ripples evolve with flux bias. (c)~Root mean square values of gain ripples for 25 different flux biases. Each of this flux is converted to a characteristic impedance $\sqrt{L_\text{J}(\Phi)/C_\text{g}}$. A local minimum is found for a small-signal impedance equals to \SI{48}{\ohm}}
\label{fig4}
\end{figure}

A useful and new feature of our STWPA is its flux tunability. It allows for the bandgap to be tuned and hence the pump frequency. \textit{In-situ} modulation of the characteristic impedance $Z_\text{c}$ is also an interesting capability to mitigate the fabrication uncertainties, mainly regarding $\bar{L}$. We now report a detailed study of sample B to illustrate such feature. A color map of its linear transmission as a function of frequency and $\Phi/\Phi_0$ is shown in~\cref{fig4}({a}). The photonic gap lies at \SI{7}{\giga\hertz} at zero flux and drops towards zero frequency for fluxes close to $|\Phi/\Phi_0|=0.5$. A jump in the gap position is visible when it goes below \SI{4}{\giga\hertz}. 

This jump is hysteretic: it occurs at a different flux whether the current is swept up or down~\cite{Pogorzalek:2017bd}.This is a characteristic signature of a non-negligible ratio $\beta$ between $\bar{L}_\text{JJ}$, the inductance of one junction of the SQUID and $L_\text{s}$, the stray inductance of the whole superconducting loop ($\beta = L_\text{s}/\bar{L}_\text{JJ}$). As shown in the inset of~\cref{fig4}({a}), by interpolating the gap position to low frequencies, we can infer the position of the crossing between the zero fluxoid domain and the $\pm 1$ fluxoid domain. Furthermore, there is a zone of coexistence between these domains. At zero frequency, the amplitude of this coexistence zone is noted $\Delta c = \Delta\Phi_\text{DC}/\Phi_\text{0}$. $\Delta c$ is directly proportional to the inductances ratio, $\Delta c = \beta/2$~\cite{lecoc2011}. From the interpolation we find $\Delta c =0.16$, and thus $\beta= 0.32$. Taking into account the kinetic inductance of the wires connecting the two Josephson junctions forming the SQUID, we find a loop inductance $L_\text{s} = \SI{21}{\pico\henry}$~\cite{planat2018understanding}. This translates into a ratio $\beta = 0.21$, which is in the same order of magnitude as the measured one. 

Our STWPA were designed to feature a characteristic impedance below \SI{50}{\ohm} at zero flux. We then expect to observe a decrease of the gain ripples down to a local sweet-spot, where the STWPA is perfectly matched to the external circuitry. Three amplification profiles taken at different flux are shown in~\cref{fig4}({b}). They all show \SI{15}{\decibel} gain for \SI{2}{\giga\hertz} bandwidth. It appears clearly that ripples are minimized for $\Phi/\Phi_0=0.3$. To quantitatively investigate impedance dependence of gain ripples, we measured gain profiles for more than 25 different magnetic fluxes from zero flux to $\Phi/\Phi_0=0.4$. Each profile was smoothed using Savitzky-Golay algorithm~\cite{Savitzky1964}. We then computed the averaged root mean square (RMS) difference between the smoothed and experimental profile on a frequency band ranging between \SI{11}{\decibel} and the maximum gain (the band corresponding to the gap is disregarded, see Appendix~\ref{app:ripples} for details). The RMS ripples value is plotted versus linear impedance in~\cref{fig4}({c}). We observe the existence of a local minimum, in a interval between \SI{45}{\ohm} and \SI{48}{\ohm}, clearly just below \SI{50}{\ohm}. This is expected as the Josephson inductance increases when pumped, and we used linear inductance to calculate flux dependent characteristic impedance. Ripples do not reach zero at the sweet-spot because of remaining impedance mismatches between the setup and the STWPA.

\section{Conclusion}
In conclusion, we demonstrated a new design of SQUID-based Traveling Wave Parametric Amplifier. This amplifier is near quantum-limited, compact and directly compatible with superconducting qubits integration. It features high gain and large bandwidth despite its simple fabrication process. The absence of resonant phase matching is compensated by engineering the dispersion and opening a photonic band-gap. A good understanding of the nonlinear behavior of this Josephson photonic crystal opens door to further more fundamental experiments with such devices~\cite{grimsmo2017squeezing}. Finally, we demonstrated the possibility to implement TWPA made from SQUID, which could lead in the future to third-order Kerr nonlinearity TWPA, based on the SNAIL geometry~\cite{Zorin2016JTWPA, Zorin2017TravelingWavePA, Frattini:2017ji, Frattini:2018ud} or to TWPA with vanishingly small fourth-order Kerr nonlinearity~\cite{Bell2015JTWPA, Zhang2017JTWPA}.

\begin{acknowledgments}
The authors would like to thank L. Del Rey, D. Dufeu, E. Eyraud and J. Jarreau for support with the experimental setup, M. Selvanayagam and D. Est\`eve for critical reading of the manuscript, P. David for help with the fabrication of the top ground and V. Milchakov for the high quality sample pictures. The samples were fabricated in the Nanofab clean room located at Institut N\'eel. This research  was supported by the ANR under contracts CLOUD (project number ANR-16-CE24-0005) and by the European Union's Horizon 2020 Research and Innovation Programme, under grant agreement No. 824109, the European Microkelvin Platform (EMP). J.P.M. acknowledges  support from the Laboratoire d\textquoteright excellence LANEF in Grenoble (ANR-10-LABX-51-01). R.D. and S.L. acknowledge support from the CFM foundation and the 'Investisements d'avenir'  (ANR-15-IDEX-02) programs of the French National Research Agency. A.R. acknowledges the European Union's Horizon 2020 research and innovation programme under the Marie Sklodowska-Curie grant agreement No 754303.
\end{acknowledgments}

\appendix
\section{Fabrication parameters}
\label{app:fabrication}

\begin{figure}
\includegraphics[width=\linewidth]{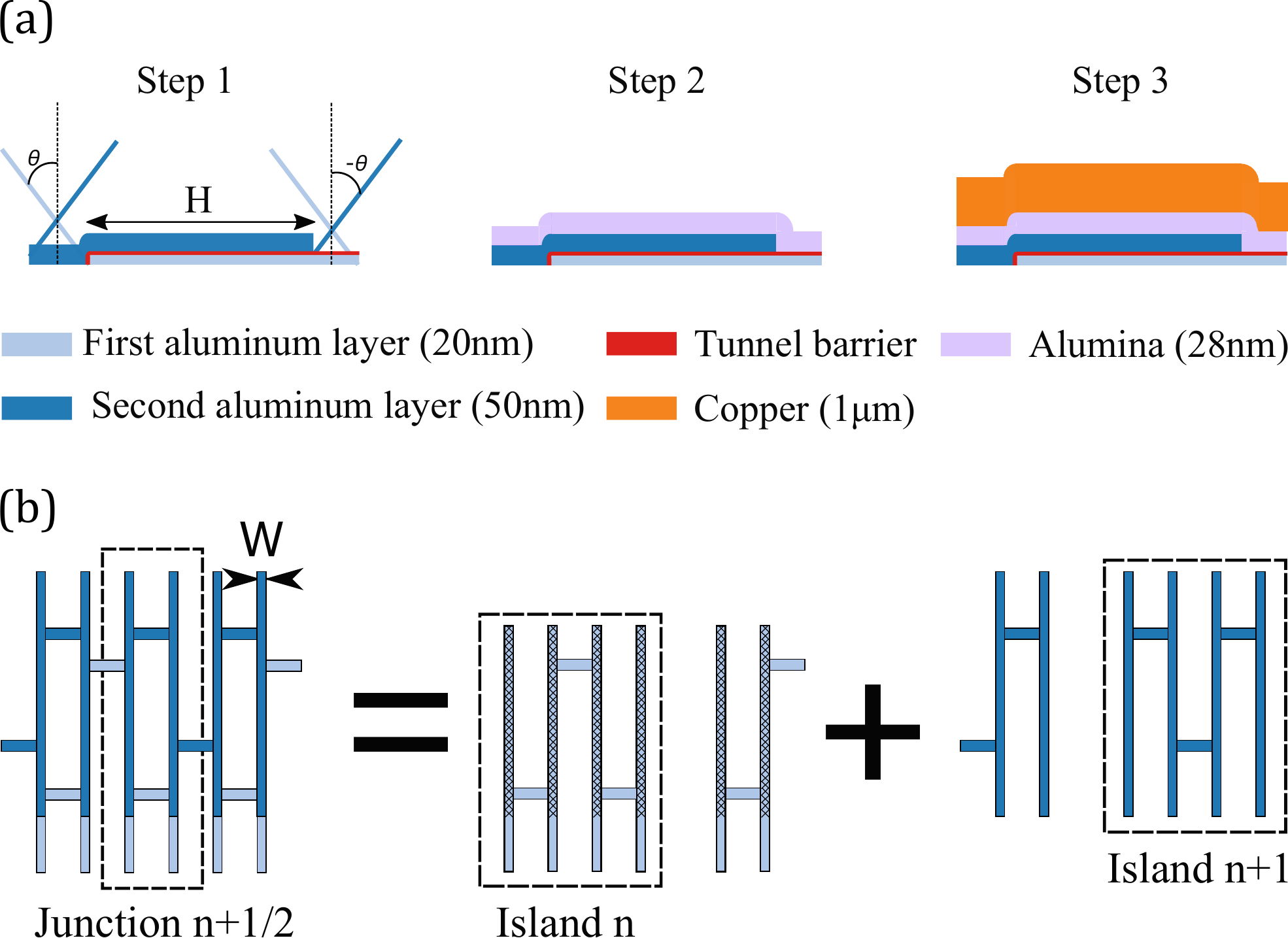}
\caption{Screening effect of the ground capacitance. (a)~Longitudinal cut of a junction during the fabrication process. The first step is a double angle evaporation of aluminum via bridge-free technique with $\theta=\SI{35}{\degree}$. The overlap between the two layers define the height H of a junction. Second step is a conformal deposition of alumina using Atomic Layer Deposition. Third step is a sputtering of a thick copper layer. The bottom layer is mostly screened by the top layer. (b)~Top view of the junctions. In this example, this short array is composed of three SQUIDs labeled $n-1/2$, $n+1/2$, $n+3/2$, and two superconducting islands, $n$ and $n+1$. Regarding the superconducting island, the capacitance between the bottom layer and the top-ground is much smaller than that between the top layer and the top-ground as the hatched region of the bottom layer is totally screened by the top layer} 
\label{parameter}
\end{figure}

The STWPA is one long array of $N_\text{J}$ SQUIDS. Each SQUID is made of two identical Josephson junctions, undergoing a sinusoidal modulation of their height~$H$. Junctions are fabricated using a bridge-free technique~\cite{lecoc2011}. This step is followed by the deposition of a \SI{28}{\nano\meter} thick conformal layer of alumina via Atomic Layer Deposition on top of the array. Finally, \SI{1}{\micro\meter} of copper is deposited on top of the alumina, working as a top metallic ground. These steps are shown schematically in~\cref{parameter}(a). The inverted strip-line geometry with very close ground allows to have a precise value of the ground capacitance~$C^\mathrm{g}$. 

In our geometry, given the extreme closeness of the ground, the capacitance between the bottom layer and the ground is much smaller than that between the top layer and the ground. Thus, the two neighboring junctions do not have the same value of $C^\mathrm{g}$. Nonetheless, the effect of this double periodicity can be discarded as it only affects the dispersion near the plasma frequency (\SI{30}{\giga\hertz}) and we work at frequencies around \SI{7}{\giga\hertz}. At such low frequencies, it is the  ground capacitance per unit length that counts, so one can use the mean value of the two islands to determine the average $\bar{C}^\text{g}=\SI{42.6}{\femto\farad}$. Furthermore, the region of the bottom layer which is not screened by the top layer, does not dependent on the junction height~$H$. Thus, it decreases the effective modulation amplitude of the ground capacitance $\zeta$.

The Josephson capacitance is taken as \SI{45}{\femto\farad} per square micrometer. The area of each junction is given by $H \times W$ (see~\cref{parameter}), and each SQUID consists of two junctions. As a result, the average SQUID has a Josephson capacitance $\bar{C}=\SI{485}{\femto\farad}$.

Since Josephson inductance is exponentially sensitive to the tunnel junction thickness, the average inverse inductance $\bar{L}^{-1}$ is the most sensitive fabrication parameter whose value cannot be reliably determined from the geometry. Thus, it was determined by fitting the dispersion relation.

\section{Theory of the Josephson photonic crystal traveling wave parametric amplifier}
\label{app:theory}

\subsection{Pump wave propagation}

We assume the pump propagation to be unaffected by signal and idler, so it can be studied separately. The solution for the pump will then determine signal and idler propagation. We assume both the modulation amplitude and the nonlinearity strength to be weak; even in this case finding the pump wave represents a non-trivial problem when the pump frequency is close to the gap. The reason is that near the gap, the solution is given by a non-perturbative mixture of two plane waves, whose coefficients can be significantly affected even by a weak nonlinearity. As a result, the gap position in frequency depends on the amplitudes of both waves. These, in turn, depend on the pump power in a non-trivial way: for a fixed incident external power, some power is reflected back, while the power that enters the chain is sensitive to the dispersion inside the chain (e.~g., the gap position), which, in turn, depends on the wave amplitudes inside the chain. Thus, the pump reflection at the chain end as well as its propagation inside the chain should be determined self-consistently for a given external incident pump power. This is quite different from the approach adopted in most studies on TWPA, where the spatial shape of the pump wave is assumed to be unaffected by the weak nonlinearity (in particular, see Ref.~\cite{Erickson:2017dy}, where the Bloch functions of a spatially modulated amplifier were taken to be power-independent).

We model the chain as a sequence of superconducting islands, labeled by an integer $n=0,1,\ldots,N$. The dynamical variables are the superconducting phases $\phi_n$ which satisfy the following equations of motion in the bulk of the chain:
\begin{align}
&\frac{d^2}{dt^2}\left[C^\mathrm{g}_n\phi_n+
C_{n+1/2}\left(\phi_n-\phi_{n+1}\right)
+C_{n-1/2}\left(\phi_n-\phi_{n-1}\right)\right]\nonumber\\
&{}+\frac{\sin(\phi_n-\phi_{n+1})}{L_{n+1/2}}
+\frac{\sin(\phi_n-\phi_{n+1})}{L_{n-1/2}}=0.
\label{eq:maineqA=}
\end{align}
Here $C^g_n$ is the capacitance between the $n$th island and the ground, while $C_{n+1/2}$ and $L_{n+1/2}$ are the capacitance and the Josephson inductance of the junction between the islands $n$ and $n+1$ (hence the half-integer index). In the following, the Josephson nonlinearity will be expanded to the leading order, $\sin\theta\approx\theta-\theta^3/6$.

The inductances and the capacitances are assumed to have a weak spatial modulation given by Eq.~(\ref{eq:modulation}). It is convenient to introduce the plasma frequency $\omega_\infty=1/\sqrt{\bar{L}\bar{C}}$ and the Coulomb screening length $\ell_s=\sqrt{\bar{C}/\bar{C}^\mathrm{g}}$.
Assuming the pump to be (i)~monochromatic, and (ii)~a mixture of at most two resonant plane waves, we look for the solution of Eq.~(\ref{eq:maineqA=}) in the form
\begin{equation}\label{eq:phipump=}
\phi_n^\mathrm{p}(t)=\left(Ae^{ikn}+Be^{ikn-iGn}\right)e^{-i\omega_\mathrm{p}t} + \mbox{c.c.},
\end{equation}
where $\omega_\mathrm{p}$ is the pump frequency, determined by the external source, while the amplitudes $A,B$ and the wave vector~$k$ should be determined from the equations; ``c.~c.'' stands for the complex conjugate. We assume $0<k<G$,  substituting this expression in Eq.~(\ref{eq:maineqA=}), omitting high spatial and temporal harmonics, and approximating $\sin(k/2)\approx{k}/2$, $\sin(G/2-k/2)\approx G/2-k/2$, $\cos(k/2)\approx\cos(G/2-k/2)\approx1$, we obtain
\begin{subequations}
\begin{widetext}\begin{align}
&\left[\frac{\omega_\mathrm{p}^2}{\omega_\infty^2\ell_s^2}
-k^2\left(1-\frac{\omega_\mathrm{p}^2}{\omega_\infty^2}\right)
+\frac{k^4}2\,|A|^2+k^2(G-k)^2|B|^2\right]A
+\left[\frac\zeta{2}\,\frac{\omega_\mathrm{p}^2}{\omega_\infty^2\ell_s^2}
+\frac\eta{2}\,k(G-k)\left(1-\frac{\omega_\mathrm{p}^2}{\omega_\infty^2}\right)\right]B=0,
\label{eq:pumpwaveA=}\\
&\left[\frac\zeta{2}\,\frac{\omega_\mathrm{p}^2}{\omega_\infty^2\ell_s^2}
+\frac\eta{2}\,k(G-k)\left(1-\frac{\omega_\mathrm{p}^2}{\omega_\infty^2}\right)\right]A
+\left[\frac{\omega_\mathrm{p}^2}{\omega_\infty^2\ell_s^2}
-(G-k)^2\left(1-\frac{\omega_\mathrm{p}^2}{\omega_\infty^2}\right)
+k^2(G-k)^2|A|^2+\frac{(G-k)^4}2\,|B|^2\right]B=0.
\label{eq:pumpwaveB=}
\end{align}\end{widetext}
From these equations, one can obtain the linear dispersion relation by neglecting $|A|^2$ and $|B|^2$ inside the square brackets and looking for~$k$ such that the obtained linear system for $A$ and $B$ admits a nonzero solution. In the nonlinear case, Eqs.~(\ref{eq:pumpwaveA=}) and~(\ref{eq:pumpwaveB=}) represent just two equations for three unknown quantities $k,A,B$. To close the equations, one should relate $A$ and $B$ to the incident power~$P_\mathrm{in}$, which represents another external control parameter, in addition to~$\omega_\mathrm{p}$.

\begin{figure}
\includegraphics[width=\linewidth]{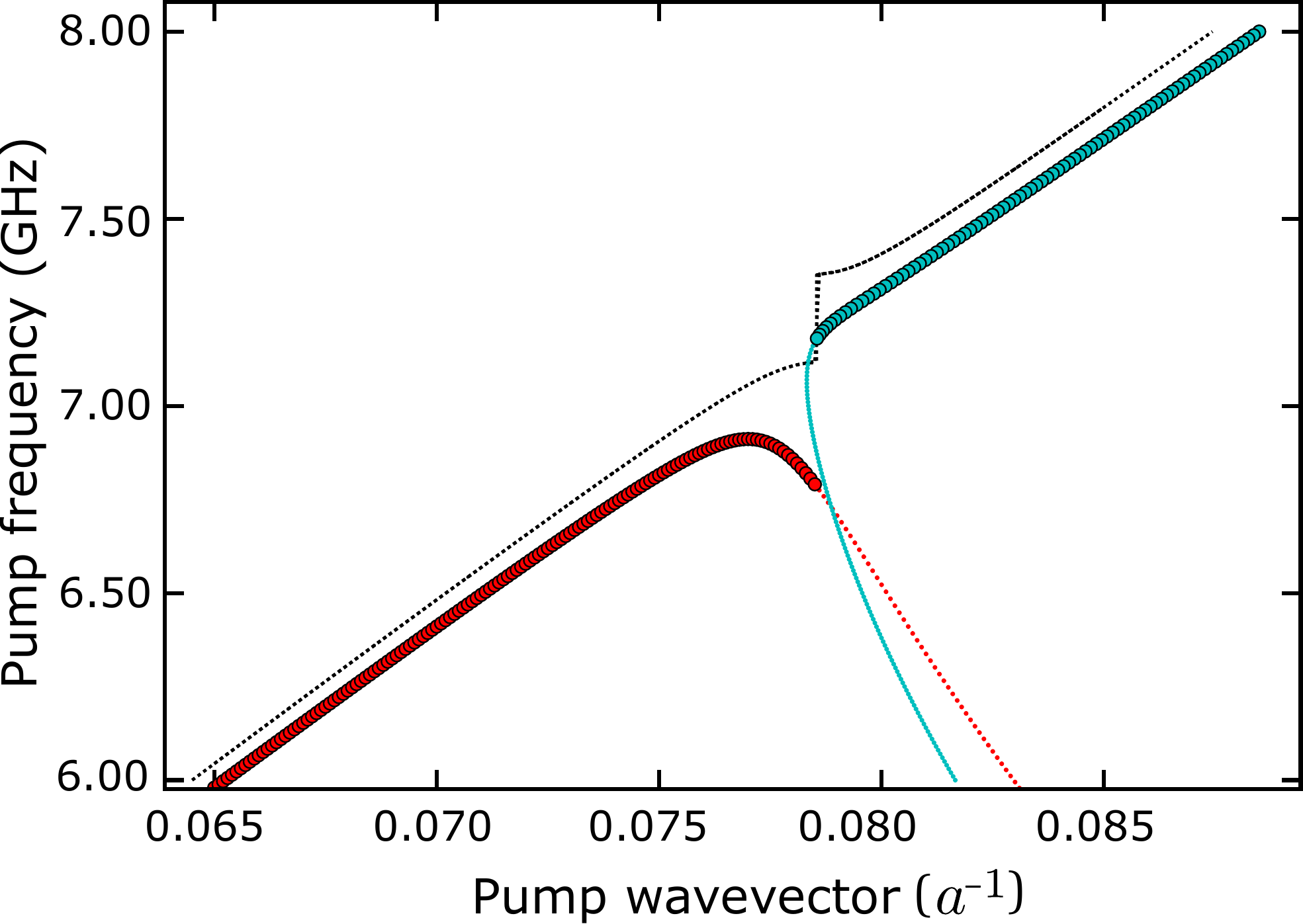}
\caption{\label{fig:PumpDisp}
Pump dispersion $\omega_\mathrm{p}$ vs $k$ in the linear case (dashed line) and at $-71\:\mbox{dBm}$ incident power (symbols) determined by looking for solutions of Eqs. (A3a)-(A3e) for each pump frequency from 6 to 8 GHz. Thick symbols indicate the physical branches with $P>0$, thin symbols -- unphysical branches with $P<0$. Note the existence of two physical solutions in a narrow interval of~$\omega_\mathrm{p}$ around 6.8~GHz. Investigation of their stability is beyond the scope of this paper.}
\end{figure}

To describe the incident microwave, we model the circuit to the left of the $n=0$ island as a linear $LC$ transmission line. Formally, this corresponds to taking the same equations~(\ref{eq:maineqA=}) and setting $C^g_n\to C_\mathrm{TL}$, $C_{n+1/2}\to0$,  $L_{n+1/2}\to L_\mathrm{TL}$, as well as linearizing $\sin(\phi_n-\phi_{n+1})\to\phi_n-\phi_{n+1}$ for all $n<0$. The transmission line impedance $Z_\mathrm{TL}$ is then given by $Z_\mathrm{TL}=\sqrt{L_\mathrm{TL}/C_\mathrm{TL}}$. The solution for $n\leq{0}$ is the sum of the incident and reflected wave, 
\[
\phi_{n<0}^\mathrm{p}(t)=\left(A_\mathrm{in}e^{ik_\mathrm{TL}n}+A_\mathrm{r}e^{-ik_\mathrm{TL}n}\right)e^{-i\omega_\mathrm{p}t}+\mathrm{c.c.},
\]
with $k_\mathrm{TL}=\omega_\mathrm{p}\sqrt{L_\mathrm{TL}C_\mathrm{TL}}$. Solution (\ref{eq:phipump=}) being valid at $n\geq{0}$, the continuity at $n=0$ requires
\begin{equation}\label{eq:continuity=}
A_\mathrm{in}+A_\mathrm{r}=A+B,
\end{equation}
while Eq.~(\ref{eq:maineqA=}) at $n=0$ gives
\begin{align}
&\frac{i\omega_\mathrm{p}{L}_{1/2}}{Z_\mathrm{TL}}\,(A_\mathrm{in}-A_\mathrm{r})
=\nonumber\\
&=\left[\left(1-\frac{\omega_\mathrm{p}^2}{\omega_\infty^2}\right)(1+\eta)
-\frac{|kA+(k-G)B|^2}2\right]\nonumber\\
&\quad{}\times[ikA+i(k-G)B].
\end{align}
The pump reflection at $n=N$ is neglected.

Finally, using the relation between the voltage~$V_n$ and the phase, $d\phi_n/dt=2eV_n/\hbar$ (with $e>0$, the electron charge being $-e$), we express the incident power $P_\mathrm{in}$ in terms of the amplitude $A_\mathrm{in}$:
\begin{equation}
P_\mathrm{in}=\frac{2\hbar^2\omega_\mathrm{p}^2}{Z_\mathrm{TL}(2e)^2}\, 
|A_\mathrm{in}|^2.
\label{eq:Pin=}
\end{equation}
\end{subequations}
We can also determine the power carried by the pump wave inside the chain. Defining it as the rate of change of all energy to the left of the site $n=0$,
\begin{align}
P={}&\frac{d}{dt}\sum_{n=1}^\infty\frac{\hbar^2}{(2e)^2}\,
\frac{C^g_n\dot\phi_n^2+C_{n+1/2}(\dot\phi_n-\dot\phi_{n+1})^2}{2}\nonumber\\
&{}+\frac{d}{dt}\sum_{n=1}^\infty\frac{\hbar^2}{(2e)^2}\,
\frac{1-\cos(\phi_n-\phi_{n+1})}{L_{n+1/2}},
\label{eq:power=}
\end{align}
and using Eq.~(\ref{eq:maineqA=}), we obtain
\begin{equation}
P=\frac{\hbar^2}{(2e)^2}\,\dot{\phi}_1
\left[C_{1/2}(\ddot{\phi}_0-\ddot{\phi}_1)
+\frac{\sin(\phi_0-\phi_1)}{L_{1/2}}\right].
\end{equation}
Inserting solution~(\ref{eq:phipump=}), we obtain.
\begin{align}
P={}&{}2\omega_\mathrm{p}{E}_J\left[(k-G/2)|A+B|^2+(G/2)(|A|^2-|B|^2)\right]
\nonumber\\ 
&{}\times\left[\left(1-\frac{\omega_\mathrm{p}^2}{\omega_\infty^2}\right)(1+\eta) -\frac{|kA+(k-G)B|^2}2\right].\label{eq:pumppower=}
\end{align}
Equations (\ref{eq:pumpwaveA=})--(\ref{eq:Pin=}) form a closed system for five unknown quantities $A,B,A_\mathrm{in},A_\mathrm{r},k$. While $A_\mathrm{in}$ and $A_\mathrm{r}$ are straightforwardly eliminated using Eqs.~(\ref{eq:continuity=}) and (\ref{eq:Pin=}), solution of the remaining three equations represents a non-trivial task. Depending on the parameters, these equations may have no solutions (corresponding to the nonlinear gap for the pump), or may have several solutions. Moreover, not all these solutions necessarily propagate in the right direction. Generally, it is not obvious \textit{a priori}, in which direction the wave (\ref{eq:phipump=}) is propagating, since it contains both positive and negative wave vectors. In this situation, one should look where the energy is flowing. Namely, for each solution, Eq.~(\ref{eq:pumppower=}) can be evaluated, and solutions corresponding to $P<0$ must be discarded. Indeed, such solutions correspond to the energy propagating in the opposite direction (we inject the pump on the left end of the chain, and want the energy to propagate from left to right). A detailed study of possible solutions for the propagating pump is beyond the scope of the present paper. Here we solve the equations iteratively by varying $\omega_\mathrm{p}$ and approaching the gap region either from higher or lower frequencies, thereby reconstructing the two branches of the power-dependent pump dispersion, as illustrated in Fig.~\ref{fig:PumpDisp}.

\begin{figure}
\includegraphics[width=\linewidth]{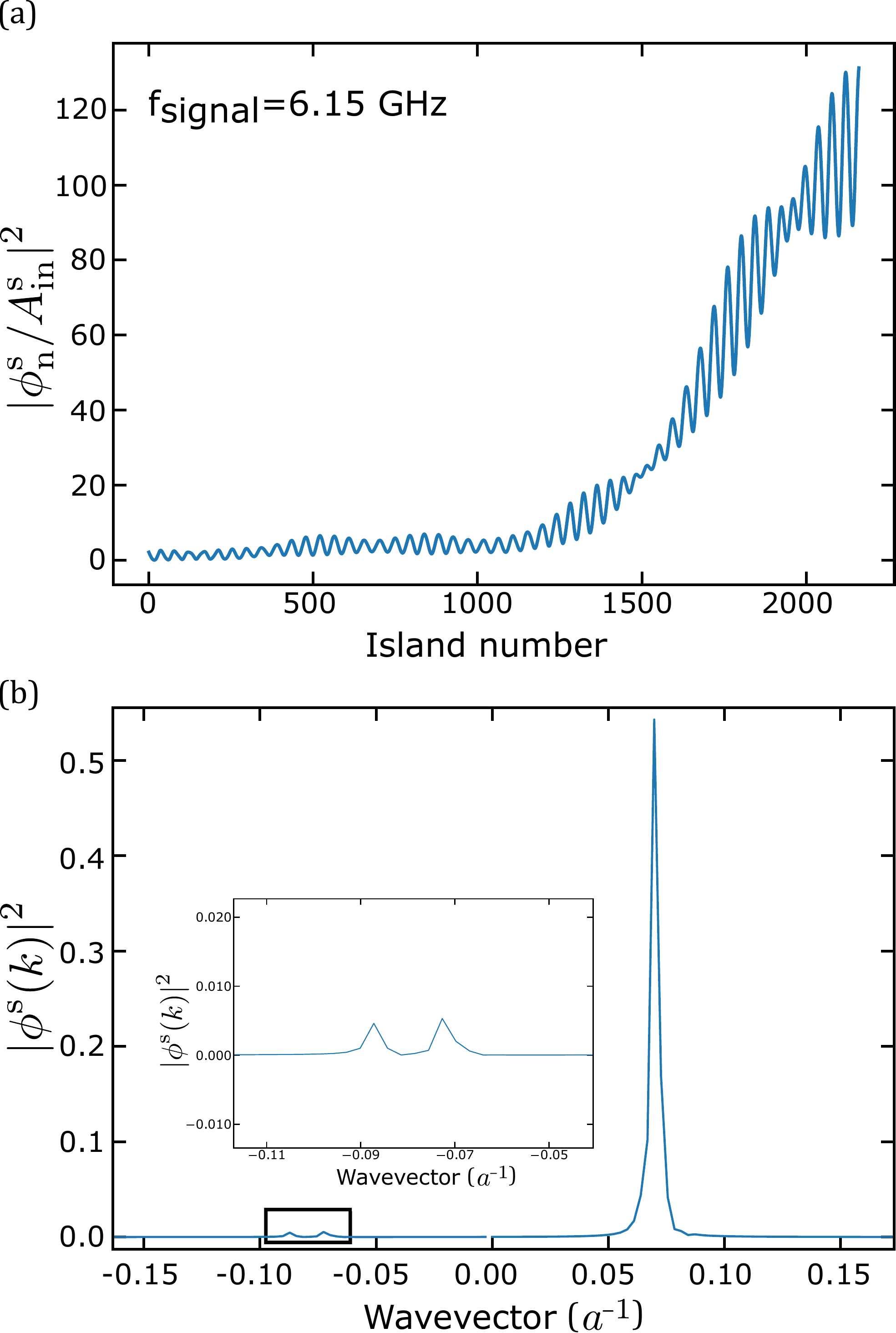}
\caption{\label{fig:phis}
(a)~$|\phi_n^\mathrm{s}|^2$ for pump frequency equals to 6.901~GHz and $-70\:\mbox{dBm}$ incident power. Signal frequency is 6.15~GHz. (b)~Fourier transform $|\phi^\mathrm{s}(k)|^2$ of the signal shown in~(a).
}
\end{figure}

\subsection{Signal and idler waves}

Having found the solution of Eq.~(\ref{eq:maineqA=}) for the pump in the form~(\ref{eq:phipump=}), we add a weak perturbation oscillating at two other frequencies, $\omega_\mathrm{s}$ (signal) and $\omega_\mathrm{i}=2\omega_\mathrm{p}-\omega_\mathrm{s}$ (idler):
\begin{equation}
\phi_n(t)=\phi_n^\mathrm{p}(t)
+\left(\phi_n^\mathrm{s}e^{-i\omega_\mathrm{s}t}+\mbox{c.c.}\right)
+\left(\phi_n^\mathrm{i}e^{-i\omega_\mathrm{i}t}+\mbox{c.c.}\right).
\end{equation}
Substituting this expression into Eq.~(\ref{eq:maineqA=}), linearizing with respect to $\phi_n^\mathrm{s}$ and $\phi_n^\mathrm{i}$, and omitting high harmonics, we obtain a closed set of linear equations for $\phi_n^\mathrm{s}$ and $\bar\phi_n^\mathrm{i}\equiv(\phi_n^\mathrm{i})^*$:
\begin{subequations}\label{eq:signalidler=}\begin{align}\label{eq:signalwave=}
&\frac{\omega_\mathrm{s}^2}{\omega_\infty^2\ell_s^2}\left(1+\zeta\cos{G}n\right)\phi_n^\mathrm{s}+J_{n+1/2}^\mathrm{s}-J_{n-1/2}^\mathrm{s}=0,\\
&\frac{\omega_\mathrm{i}^2}{\omega_\infty^2\ell_s^2}\left(1+\zeta\cos{G}n\right)\bar\phi_n^\mathrm{i}+\bar{J}_{n+1/2}^\mathrm{i}-\bar{J}_{n-1/2}^\mathrm{i}=0,
\label{eq:idlerwave=}
\end{align}\end{subequations}
where we introduced compact notations
\begin{subequations}\begin{align}\label{eq:Jnus=}
J_\nu^\mathrm{s}\equiv{}&{}\left(1+\eta\cos{G}\nu\right)\nonumber\\
&\times\left[\left(1-\frac{\omega_\mathrm{s}^2}{\omega_\infty^2}-|\partial\phi^\mathrm{p}_\nu|^2\right)\partial\phi^\mathrm{s}_\nu
-\frac{1}2\left(\partial\phi^\mathrm{p}_\nu\right)^2\partial\bar\phi^\mathrm{i}_\nu\right],\\
\bar{J}_\nu^\mathrm{i}\equiv{}&{}\left(1+\eta\cos{G}\nu\right)\nonumber\\
&\times\left[\left(1-\frac{\omega_\mathrm{i}^2}{\omega_\infty^2}-|\partial\phi^\mathrm{p}_\nu|^2\right)\partial\bar\phi^\mathrm{i}_\nu
-\frac{1}2\left((\partial\phi^\mathrm{p}_\nu)^*\right)^2\partial\phi^\mathrm{s}_\nu\right],\label{eq:Jnui=}
\end{align}\end{subequations}
and denoted $\partial\phi_\nu\equiv\phi_{\nu+1/2}-\phi_{\nu-1/2}$. These expressions with $\nu=1/2,\,3/2,\,\ldots,\,N-1/2$ define equations (\ref{eq:signalwave=}), (\ref{eq:idlerwave=}) for $n=1,2,\ldots,N-1$. To close the system, we need two equations for $n=0$ and two more for $n=N$. Note that in the description of the pump wave no boundary condition at $n=N$ was invoked; this was because the pump reflection was neglected, so the propagation was the same as in a semi-infinite chain. For the signal and idler waves we want to include the reflection; even though weak due to impedance matching, the reflected waves might be amplified, and we want to account for this process.

\begin{figure}
\includegraphics[width=\linewidth]{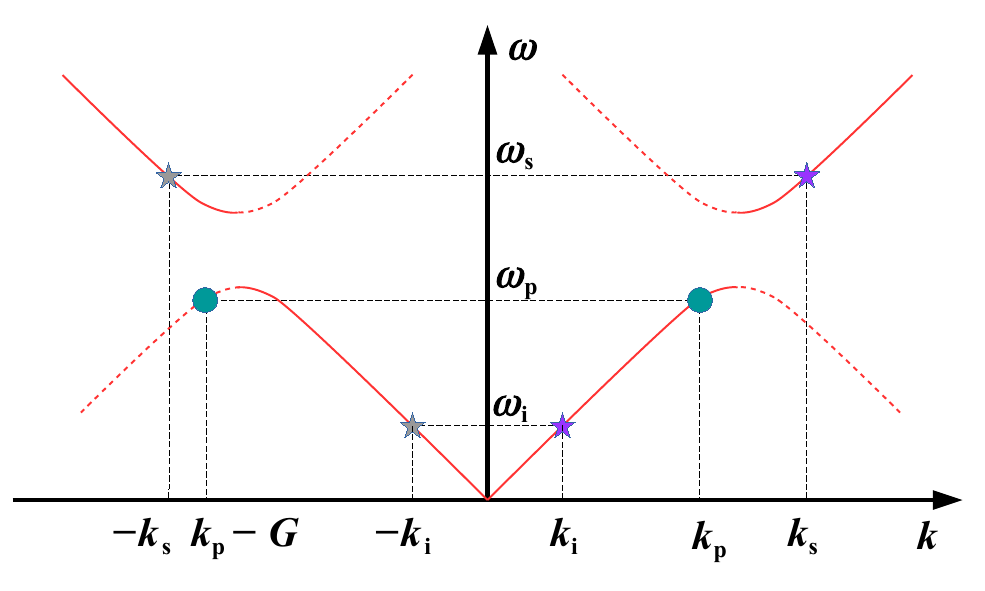}
\caption{\label{fig:AmplifDisp}
A schematic view of the gapped dispersion (red curves, the dashed ones corresponding to folded dispersion) in the extended Brillouin zone picture. The green circles correspond to the two components of the pump wave with wave vectors $k_\mathrm{p}>0$ and $k_\mathrm{p}-G<0$, corresponding to the frequency $\omega_\mathrm{p}$ near the gap edge. The purple stars represent signal and idler wave vectors $k_\mathrm{s},k_\mathrm{i}>0$, corresponding to the frequencies $\omega_\mathrm{s},\omega_\mathrm{i}$, while the grey stars represent $-k_\mathrm{s},-k_\mathrm{i}$, corresponding to the same frequencies. For $k_\mathrm{p}\approx{G}/2$, phase matching of $k_\mathrm{s},k_\mathrm{i}$ with $2k_\mathrm{p}$ may automatically yield phase matching of $-k_\mathrm{s},-k_\mathrm{i}$ with $2(k_\mathrm{p}-G)$ leading to potentially dangerous backward amplification.
}
\end{figure}

As before, we model the circuit to the left of $n=0$ and to the right of $n=N$ as a linear $LC$ transmission line with the same parameters $L_\mathrm{TL}$, $C_\mathrm{TL}$. We assume that on the left we have a given incoming signal with amplitude~$A_\mathrm{in}^\mathrm{s}$, an unknown outgoing signal with amplitude $A_\mathrm{in}^\mathrm{s}-\phi_0^\mathrm{s}$ (as follows from the continuity at $n=0$), zero incoming idler, and unknown outgoing idler with amplitude $\phi_0^\mathrm{i}$, while on the right we have only outgoing signal and idler with unknown amplitudes $\phi_N^\mathrm{s,i}$.
This amounts to defining $J_{-1/2}^\mathrm{s}$ and $J_{N+1/2}^\mathrm{s,i}$ in Eqs.~(\ref{eq:signalwave=}) and~(\ref{eq:idlerwave=}) for $n=0,N$ as
\begin{subequations}\begin{align}
&J_{-1/2}^\mathrm{s}\to
\frac{\partial\phi_{-1/2}^\mathrm{s}}{L_\mathrm{TL}\omega_\infty^2\ell_s^2C_g}
=\frac{i\omega_\mathrm{s}}{\omega_\infty\ell_s}\,\frac{Z_g}{Z_\mathrm{TL}}
\left[2A_\mathrm{in}^\mathrm{s}-\phi_0^\mathrm{s}\right],\\
&J_{-1/2}^\mathrm{i}\to
\frac{\partial\bar\phi_{-1/2}^\mathrm{i}}{L_\mathrm{TL}\omega_\infty^2\ell_s^2C_g}
=\frac{i\omega_\mathrm{i}}{\omega_\infty\ell_s}\,\frac{Z_g}{Z_\mathrm{TL}}\,\bar\phi_0^\mathrm{i},\\
&J_{N+1/2}^\mathrm{s}\to
\frac{\partial\phi_{N+1/2}^\mathrm{s}}{L_\mathrm{TL}\omega_\infty^2\ell_s^2C_g}
=\frac{i\omega_\mathrm{s}}{\omega_\infty\ell_s}\,\frac{Z_g}{Z_\mathrm{TL}}\,\phi_N^\mathrm{s},\\
&J_{N+1/2}^\mathrm{i}\to
\frac{\partial\bar\phi_{N+1/2}^\mathrm{i}}{L_\mathrm{TL}\omega_\infty^2\ell_s^2C_g}
=-\frac{i\omega_\mathrm{i}}{\omega_\infty\ell_s}\,\frac{Z_g}{Z_\mathrm{TL}}\,\bar\phi_N^\mathrm{i},
\end{align}\end{subequations}
where we defined the impedances $Z_\mathrm{TL}\equiv\sqrt{L_\mathrm{TL}/C_\mathrm{TL}}$, $Z_g\equiv1/(\omega_\infty\ell_sC_g)$ of the transmission line and of the chain, respectively. Now the linear system (\ref{eq:signalwave=}), (\ref{eq:idlerwave=}) is closed, so all $\phi_n^\mathrm{s,i}$ can be found by the brute force numerical solution of this system. The resulting amplitudes are all proportional to $A_\mathrm{in}^\mathrm{s}$, so the ratios $\phi_N^\mathrm{s}/A_\mathrm{in}^\mathrm{s}$ and $(A_\mathrm{in}^\mathrm{s}-\phi_0^\mathrm{s})/A_\mathrm{in}^\mathrm{s}$ give the signal amplification in transmission and reflection, respectively.

\begin{figure}
\includegraphics[width=\linewidth]{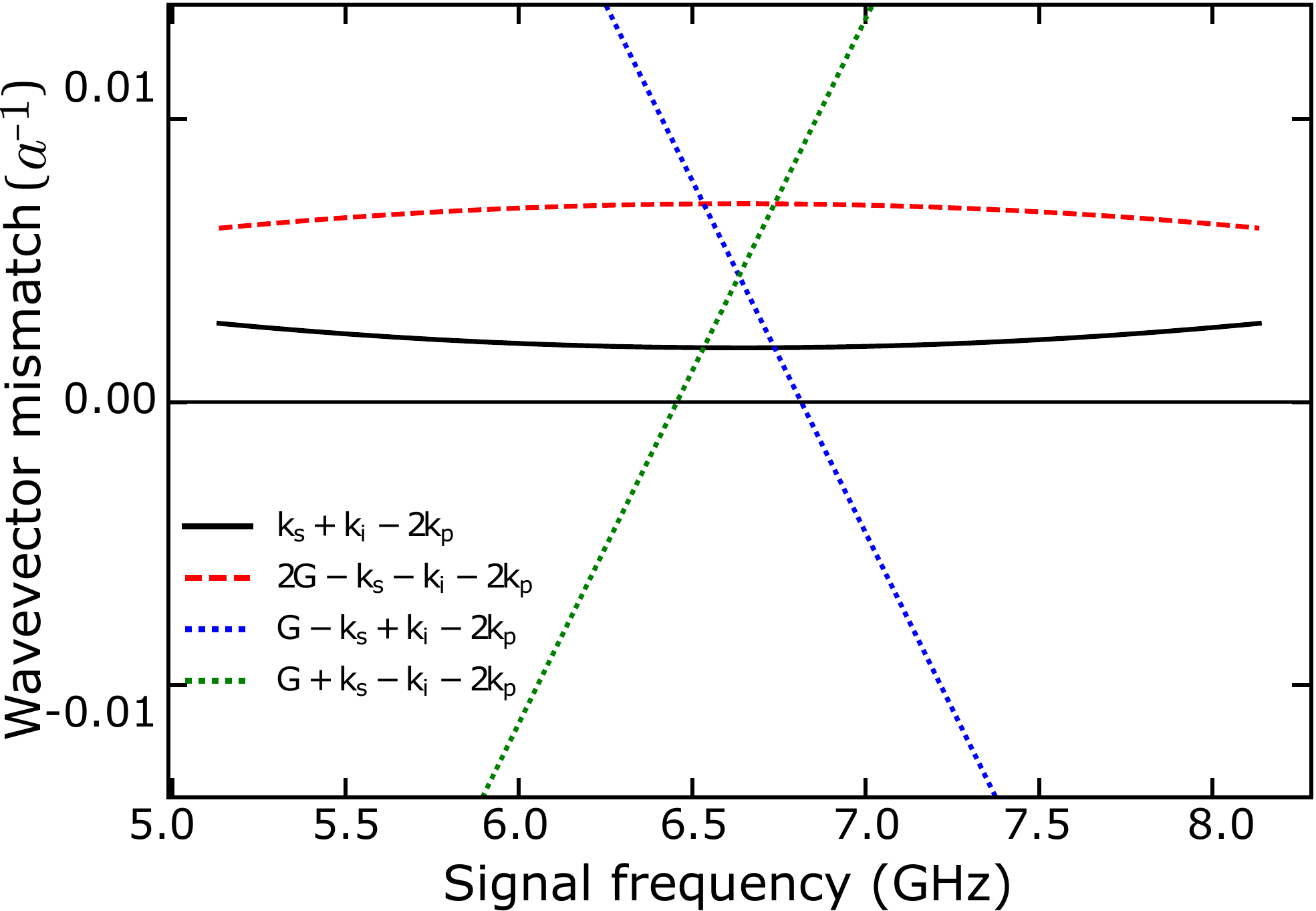}
\caption{\label{fig:Mismatch}
Wave vector mismatch for four wave combinations -- candidates for amplification. The solid line corresponds to the desired amplifier operation. The dashed line corresponds to the potential parasitic amplification of backward propagating signal and idler. The two dotted lines correspond to signal and idler propagating in opposite directions. Pump frequency and power are 6.901~GHz and $-70\:\mbox{dBm}$, respectively.
}
\end{figure}

The described procedure is very general and does not make any \emph{a priori} assumptions on which waves are amplified; waves propagating forward, or reflected back, or even evanescent (if the signal or idler frequency happens to be in the gap) are all treated on equal footing. To gain some qualitative insight into the amplification process, one should analyse the structure of the solution for $\phi_n^\mathrm{s}$. In Fig.~\ref{fig:phis}(a) we plot $|\phi_n^\mathrm{s}|^2$, which shows the exponential growth in the forward direction, corresponding to the amplification.  The wave content of the obtained solution can be better inferred via its spatial Fourier transform,
\begin{equation}\label{eq:phiK=}
\phi^\mathrm{s}(k)=\sum_{n=0}^N\phi^\mathrm{s}_ne^{-ikn}.
\end{equation}
We plot $|\phi^\mathrm{s}(k)|^2$ Fig.~\ref{fig:phis}(b). It shows three peaks, corresponding to the three components of the signal wave at $k_\mathrm{s}>0$, $-k_\mathrm{s}<0$ and $k_\mathrm{s}-G<0$. The $-k_\mathrm{s}$ component corresponds to the reflected wave. 

It has been pointed out~\cite{Obrien2014resonant} that a TWPA with periodic spatial modulation is potentially subject to the parasitic effect of backward parametric amplification. Namely, if the pump wave has the profile $\phi^\mathrm{p}_n=A_\mathrm{p}e^{ik_\mathrm{p}n}+B_\mathrm{p}e^{i(k_\mathrm{p}-G)n}$, and the amplification of the signal and idler waves with $k_\mathrm{s},k_\mathrm{i}>0$ is efficient due to phase matching with the $k_\mathrm{p}\approx{G}/2$ component of the pump, $2k_\mathrm{p}\approx k_\mathrm{s}+k_\mathrm{i}$, this may automatically yield phase matching of $-k_\mathrm{s},-k_\mathrm{i}$ with the $k_\mathrm{p}-G$ component of the same pump, $-k_\mathrm{s}-k_\mathrm{i}\approx2(k_\mathrm{p}-G)$, as shown schematically in Fig.~\ref{fig:AmplifDisp}.

To study possible phase matched combinations, let us assume $\omega_\mathrm{s},\omega_\mathrm{i}$ to be sufficiently far from the gap, so that one can straightforwardly define the signal and idler dispersion, corrected by the pump:
\begin{equation}\label{eq:ksi=}
k_\mathrm{s,i}^2=\frac{\omega_\mathrm{s,i}^2/(\omega_\infty\ell_s)^2}%
{1-\omega_\mathrm{s,i}^2/\omega_\infty^2-k_p^2|A_p|^2-(k_p-G)^2|B_p|^2}.
\end{equation}
This expression follows from Eqs.~(\ref{eq:signalidler=}) when terms proportional to $e^{i(k_\mathrm{s}-G)n}$, $e^{i(k_\mathrm{i}-G)n}$ are neglected. Wave vector mismatch for several combinations is plotted in Fig.~\ref{fig:Mismatch}. (When all frequencies are near the gap, it is not possible to define a simple expression for wave vectors and a single combination  characterizing the phase mismatch, so the simple picture of Fig.~\ref{fig:Mismatch} is not valid; still, the bandwidth of our TWPA well exceeds the gap width, so the assumption of $\omega_\mathrm{s},\omega_\mathrm{i}$ being far from the gap is justified in most of the frequency interval.) From the figure it is seen that the standard forward amplification process has the smallest phase mismatch, while for the parasitic backward amplification it is significantly higher. This happens because the detuning of the pump wave vector $k_\mathrm{p}$ from $G/2$ in our experiment, although small, ($|k_\mathrm{p}$ - $G/2|=0.013$), while $G/2=0.0785$ is sufficient to suppress the phase matching for the backward amplification. This mismatch can also be understood graphically using~\cref{fig:AmplifDisp}. The forward process involves the green circle and the two purples stars. The three of them are well aligned. On the other hand the backward amplification is described by the green circle and the grey stars. Their misalignment appears clearly.

\section{Experimental setup}
\label{app:setup}

\begin{figure}
\includegraphics[width=\linewidth]{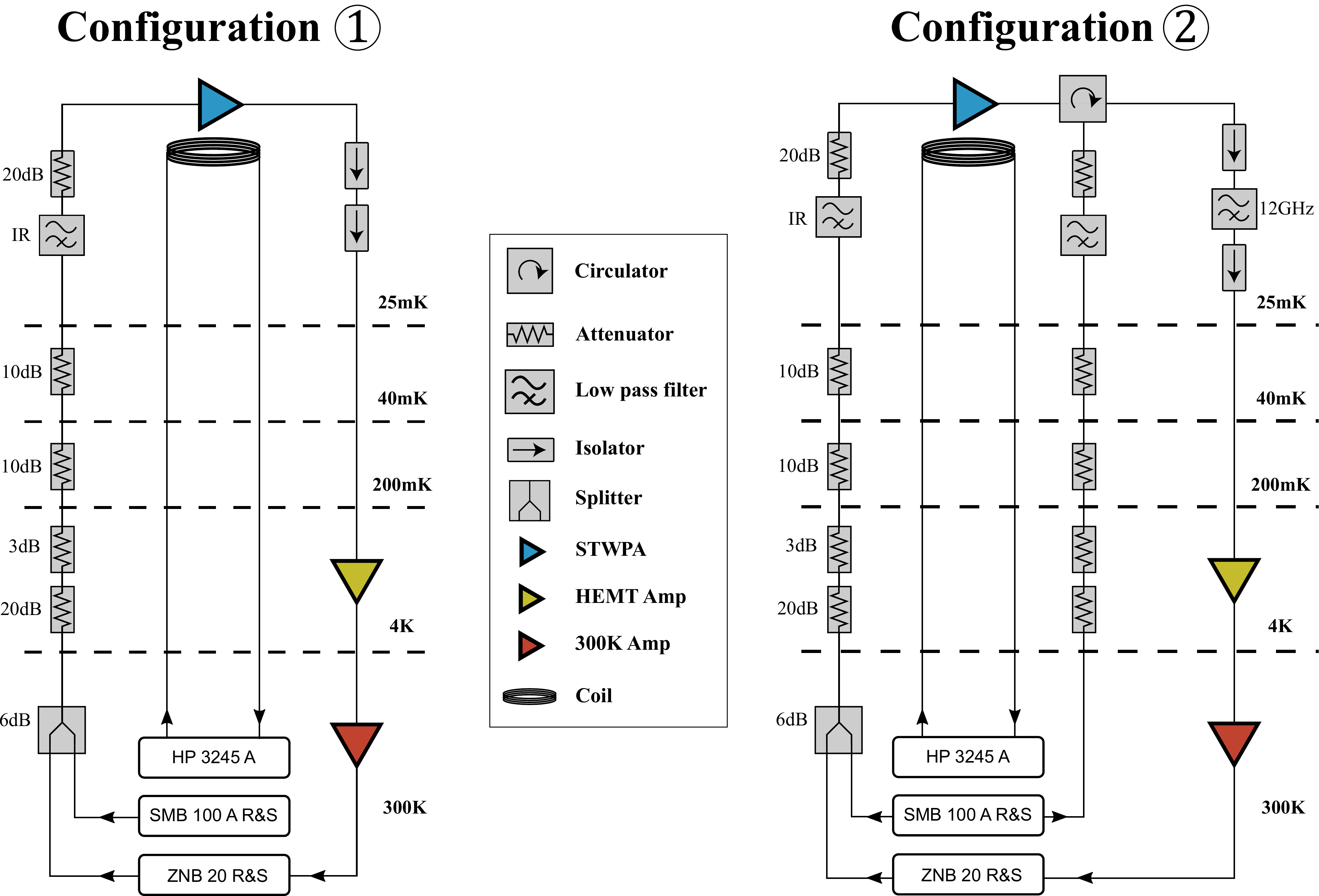}
\caption{ Measurement setup. Configuration \ding{192} is a standard microwave transmission measurement setup. Configuration \ding{193} has one more input line. This configuration can be used for the same purpose as \ding{192} (pump and signal propagate in the same direction), or to probe backward gain (pump and signal propagate in opposite directions). Attenuators and filter of the leftmost and middle RF lines are nominally identical.} 
\label{meas_setup}
\end{figure}

The samples were measured with a Vector Network Analyzer (model ZNB20 from Rohde \& Schwarz) in a dilution refrigerator with a base temperature of \SI{25}{\milli\kelvin}. An additional microwave source (Rhodes \& Schwarz SMB 100 A) was used as a pump, while a global magnetic field was applied via an external superconducting coil powered by DC current source (HP 3245 A). Both the coil and the sample were held inside a mu-metal magnetic shield. There are two measurement configurations: the first configuration, labeled \ding{192}, is a standard configuration where the STWPA is directly followed by two isolators and a commercial HEMT amplifier (model LNF-LNC1-12A from Low Noise Factory), whereas the second configuration, labeled \ding{193}, has a microwave circulator (\SI{4}{\giga\hertz}-\SI{8}{\giga\hertz}) between the output of the STWPA and the isolators to probe the backward gain, i.e. the amplification of any signal counter-propagating with respect to the pump direction (see ~\cref{meas_setup}). Configuration \ding{193} can be used to measure both in a standard way (signal and pump propagating in the same direction) and backward gain (opposite directions). Experimental data shown on~\cref{fig2}, \cref{fig3}({b}) and~\cref{fig3}({c}) below were taken with configuration \ding{193}. The rest was taken with configuration \ding{192}. 

Attenuation and phase calibrations of the setup have been done by measuring a dummy \SI{50}{\ohm} CPW transmission line prior to the STWPA. This setup calibration protocol requires two cooldowns but is expected to be as precise as using an \textit{in-situ} microwave switch, since we have noticed that the attenuation in the lines differs by less than \SI{0.5}{\decibel} from one cool down to an other.

\section{Input line calibration}
\label{app:input}

\begin{figure}
\includegraphics[width=\linewidth]{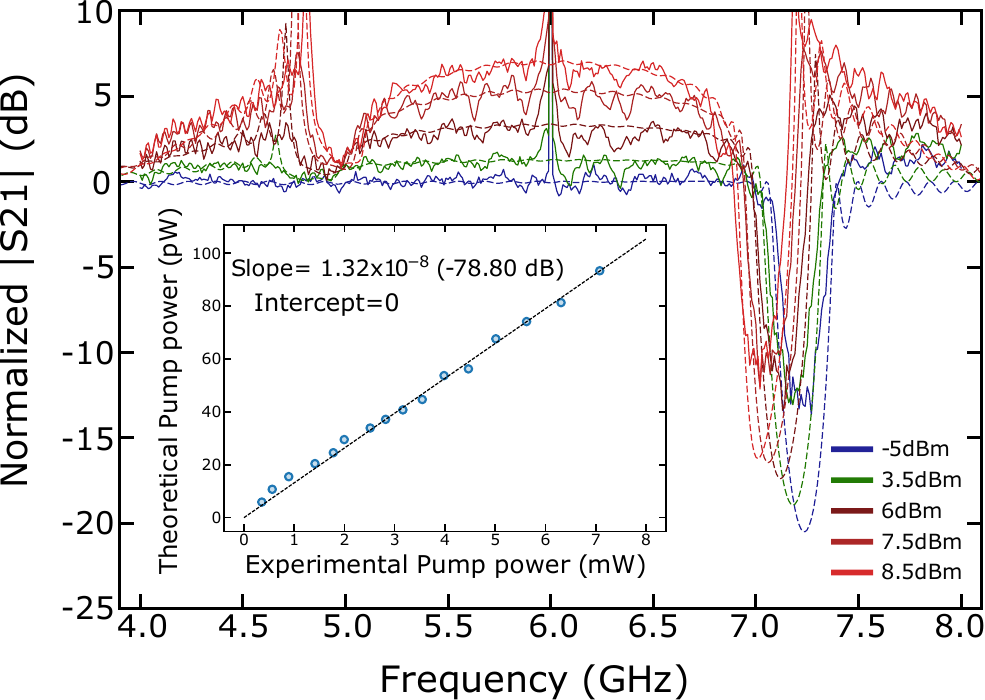}
\caption{Calibration via cross-Kerr shift. Comparison between experiment (full lines) and theory (dashed lines) for a pump tone sent at \SI{6}{\giga\hertz} at various power. This figure is similar to Fig.~\ref{fig2}, but with more pump power, demonstrating the robustness of the model. Inset: experimental and theoretical pump power needed to get the same frequency shift of the gap. The linear fit gives a system attenuation  $A_\text{total}=\SI{-78.8}{\decibel}$.} 
\label{calibration}
\end{figure}

Calibration of the input line has been carried out by taking advantage of the power dependent cross-Kerr shift of the gap frequency position. We calibrated precisely the input line at \SI{6}{\giga\hertz}, since we calibrated the added noise of the STWPA between \SI{5.995}{\giga\hertz} and \SI{6.145}{\giga\hertz}. As shown in Sec.~\ref{sec:transmission}, by sending a powerful pump tone at \SI{6}{\giga\hertz}, we observe a red shift due to Kerr effect, as predicted by the theory. Since all the array parameters are set during the linear response study (fit of the linear dispersion relation), the only remaining free parameter to get an agreement between theory and experience is the pump power. Thus, by comparing the pump power needed to shift the gap by $\Delta f$ experimentally and theoretically, we can in principle infer the attenuation between the output of the pump source and the input of the STWPA. As shown in~\cref{calibration}, by keeping the attenuation between the experimental and theoretical pump powers constant to \SI{-78.8}{\decibel}, we manage to reproduce very well the maximum gain, bandwidth and frequency position of the gap. In the inset, we report the pump power needed to obtain the same gap frequency shift. We obtain a linear relation between the 'hot' and 'cold' pump power. The slope gives the total attenuation $A_\text{total}=\SI{-78.8}{\decibel}$ 

As explained in Sec.~\ref{sec:transmission}, this total attenuation takes into account the attenuation of the system $A_\text{system}$ and the attenuation due to the STWPA itself $A_\text{twpa}$. In order to discriminate them, we have $A_\text{twpa}=e\exp(k''_\text{p}N_\mathrm{J}a/{2})$, where $k''_\text{p}$ is the imaginary part of the pump wavevector calculated from our theory, $N_\mathrm{J}a$ is the total length of the array and $\tan{\delta}$ is the high power loss tangent, inferred from experimental results. We find $A_\text{twpa}=\SI{-1.80}{\decibel}$. 

\section{Noise calibration}
\label{app:noise}

\begin{figure}
\includegraphics[width=\linewidth]{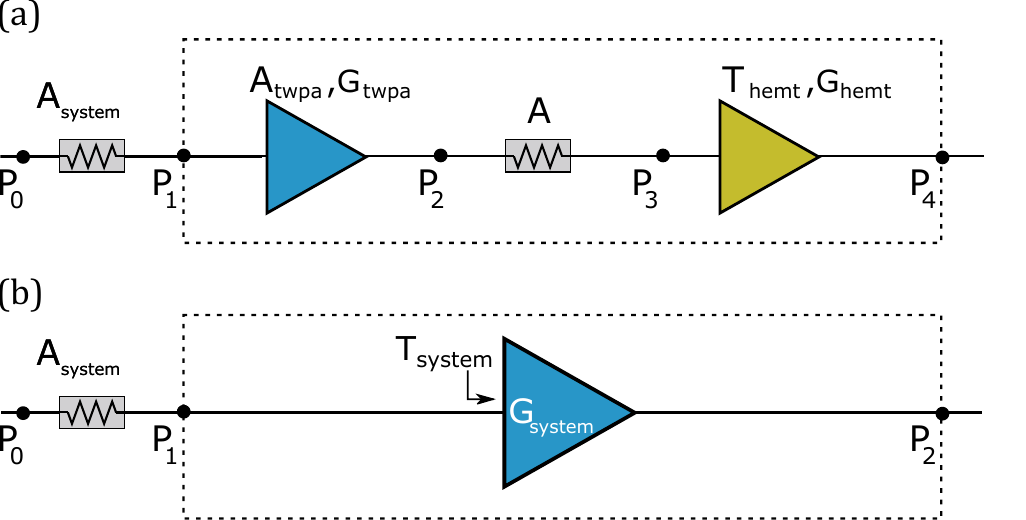}
\caption{Schematic representation of the system in terms of attenuators and amplifiers. (a)~The system attenuation is $A_\text{system}$. Inside the dotted box, there is the STWPA, modelled as an attenuator combined with a lossless amplifier, followed by an unknown attenuation $A$ and the rest of the chain (HEMT and room temperature amplifier), whose input noise is dominated by the HEMT noise. (b)~Simpler modelling of the system as a single amplifier $G_\text{system}$, whose input noise $T_\text{system}$ depends on whether or not the STWPA pump is turned on.} 
\label{system_schema}
\end{figure}

\begin{figure}
\includegraphics[width=\linewidth]{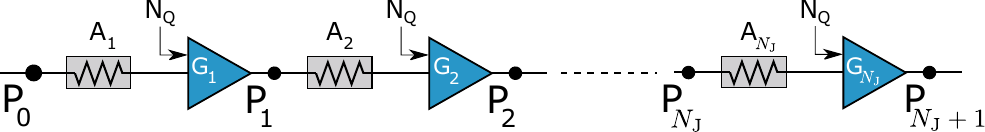}
\caption{Modeling of the STWPA as a cascade of attenuators and amplifiers. Every attenuator $A_\text{i}$ is equals to $A_\text{twpa,lp}/N_\text{J}$ and every amplifier $G_\text{i}$ is equals to $G_\text{twpa}/N_\text{J}$. We make the assumption that every amplifier is quantum limited with an input noise $N_\text{Q}=1/2$.} 
\label{TWPA_cascade}
\end{figure}

In this section, we explicit the methodology followed to get the input noise of the system composed of the STWPA, the HEMT amplifier and the room temperature amplifier. The reference plane for this system noise characterisation is defined at the input of the STWPA, point  $\text{P}_\text{1}$. We made the assumption that the noise at the system input $\text{P}_\text{1}$ [in~\cref{TWPA_cascade}({a})] is only made of quantum noise $\text{N}_\text{Q}$. We measure a PSD equals to the noise of the whole system by using a spectrum analyser plugged at the system output $\text{P}_\text{4}$ [in~\cref{system_schema}({a})]. This system can be modelled as a single amplifier $G_\text{system}$ with an input noise $T_\text{system}$ [see~\cref{system_schema}({b})]. The total system gain $G_\text{system}$ is known by measuring the full transmission $|S_{21}|^{2}$ (at very low power), and subtracting the system attenuation $A_\text{system}$ from it (see~\cref{system_schema}). Whether the STWPA pump is on or off [blue and green lines respectively, in Fig.~\ref{fig3}(c)], we obtain a system noise close to the quantum limit or a system noise around $T'_\text{hemt}=\SI{10}{\kelvin}$. Two important points must be stressed. First, the system noise when the STWPA pump is off is higher than just the HEMT noise temperature (\SI{3}{\kelvin}) because of the attenuation between the HEMT and the system input, $A_\text{twpa}$ and A. Second, the system noise does not reach the quantum limit when the STWPA is pumped. This is the consequence of losses within the SWTPA and finite gain of the STWPA compared to the effective HEMT noise. 

We first calculated what would be the STWPA noise itself. We followed Macklin \textit{et al.}~\cite{macklin2015near} and modeled the STWPA as a cascade of $N_\text{J}$ cells made of an attenuator and an amplifier. Each attenuator is given by $A_\text{i}=A_\text{twpa,lp}/N_\text{J}$ where $A_\text{twpa,lp}=\SI{-5}{\decibel}$ is the attenuation within the STWPA for very low power, and each amplifier has a gain $G_\text{i}=G_\text{twpa}/N_\text{J}$ (see~\cref{TWPA_cascade}). In this model, every amplifier is quantum limited with an added noise of half a quanta $N_\text{Q}$ and we consider that the input noise of the first unit cell is quantum limited ($N_\text{0}=N_\text{Q}$). At each point $\text{P}_\text{i}$, the noise in terms of added quanta is: 
\begin{equation}\label{eq:Ntwpa}
N_\text{i}=N_\text{i-1}A_\text{i}G_\text{i} + N_\text{Q}(1-A_\text{i})G_\text{i} + N_\text{Q}(G_\text{i}-1)
\end{equation}
Finally, the output STWPA noise at point $\text{P}_{N_\text{J}+1}$ can be normalized by the total transmission of the STWPA $G_\text{twpa} \times A_\text{twpa}$ in order to get the added noise by the STWPA alone [brown dotted line in Fig.~\ref{fig3}(c)]. To take into account the second channel of noise, namely the effective HEMT noise temperature, we simply considered the total system noise as being $T_\text{system}=T_\text{twpa} + T'_\text{hemt}/G_\text{twpa}$, giving the black dotted line in Fig.~\ref{fig3}(c). The good agreement with experimental results shows that we identified correctly the two main channels of added noise: dielectric loss within the STWPA and finite gain compared to the effective HEMT noise temperature. 

\section{SNR improvement and gain compression}
\label{app:SNR}
 
\begin{figure}
\includegraphics[width=\linewidth]{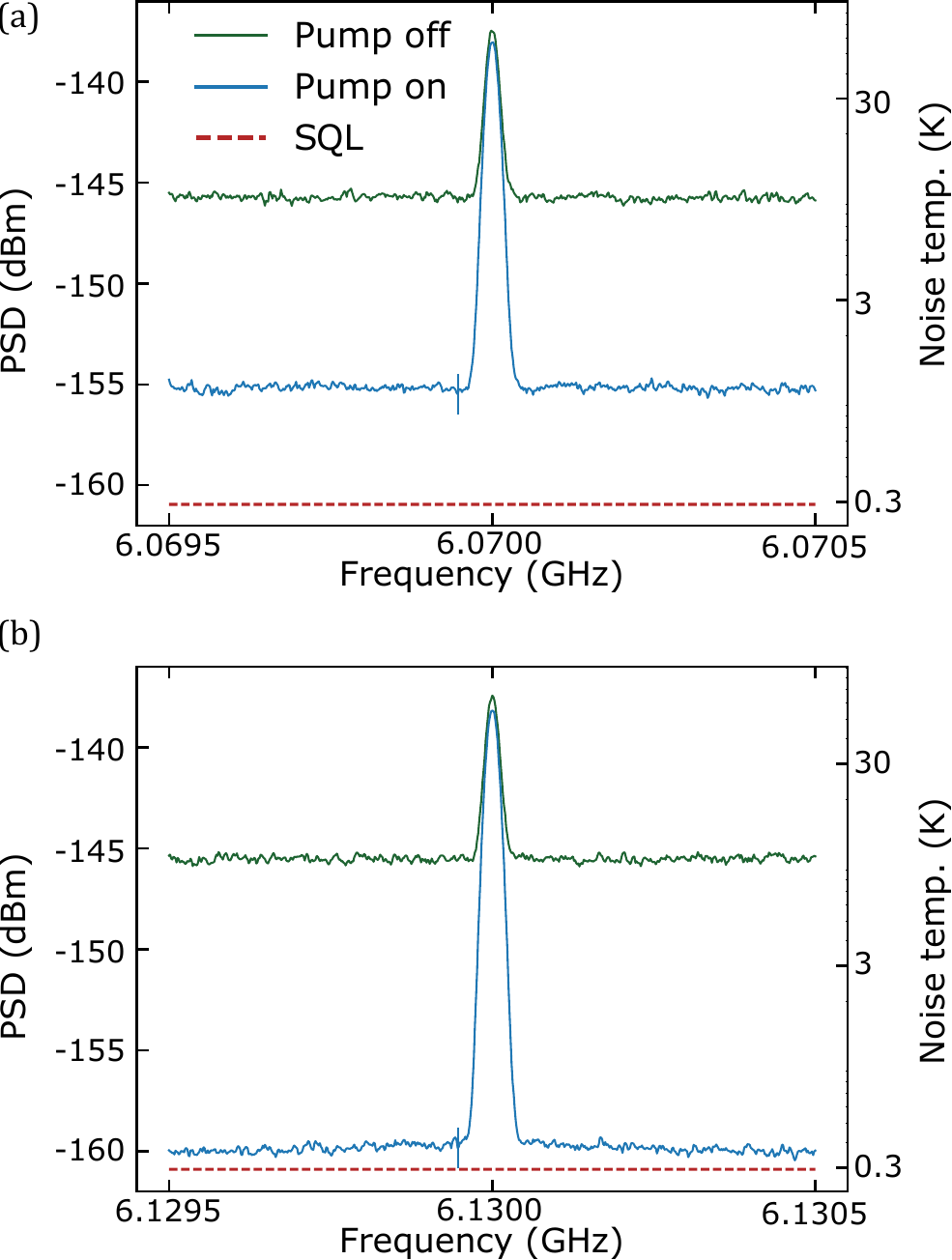}
\caption{ SNR improvement. PSD normalised by the total system gain. An improvement of up to \SI{15}{\deci\belmilliwatt} is observed. Gain ripples lead to different SNR whether the signal frequency in on a gain peak or dip. (a) SNR improvement measured on a gain dip. (b) SNR improvement measured on a gain peak.} 
\label{SNR}
\end{figure}

In this section, we show the SNR improvement of a signal sent at very low power (\SI{-135}{\deci\belmilliwatt}) and frequency corresponding to a peak and a dip in the STWPA gain (\SI{6.1300}{\giga\hertz} and \SI{6.0700}{\giga\hertz}, respectively). We plot on the same figure (see~\cref{SNR}) the PSD read with the spectral analyzer when the STWPA pump is on and off and normalise it by the total system gain (see~\cref{system_schema}). We observe a SNR improvement of \SI{15}{\deci\belmilliwatt} at a gain peak and \SI{10}{\deci\belmilliwatt} on a gain dip.

\begin{figure}
\includegraphics[width=\linewidth]{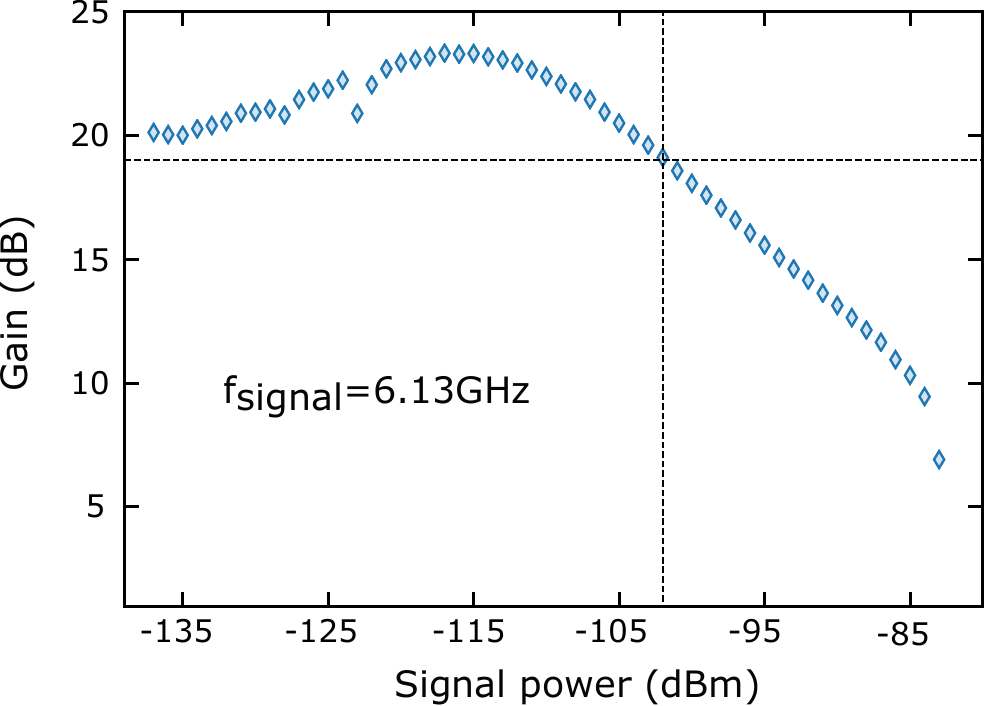}
\caption{Gain compression. Measured gain as a function of the signal power for $f_\text{signal}=\SI{6.13}{\giga\hertz}$. The gain is compressed by \SI{1}{\decibel} when the input signal power is equal to \SI{-102}{\deci\belmilliwatt}.} 
\label{1dbcp}
\end{figure}

We also investigate the power saturation of the same STWPA, in the same configuration. Knowing the attenuation of the input line, we can infer the power required to compress the gain by \SI{1}{\decibel}. In~\cref{1dbcp}, we plot the gain at a constant frequency $f_\text{signal}=\SI{6.13}{\giga\hertz}$. We observe a gain increase at intermediate signal power and a drop for higher signal power. The gain rise is caused by the power dependance of the losses within the STWPA. When the signal power increases, it suffers from less losses, which results in a higher gain. For higher signal power, the gain starts to drop. We measure a \SI{1}{\decibel} compression point of \SI{-102}{\deci\belmilliwatt} at \SI{20}{\decibel} gain (see~\cref{1dbcp}).

\section{RMS value of gain ripples}
\label{app:ripples}

\begin{figure}
\includegraphics[width=\linewidth]{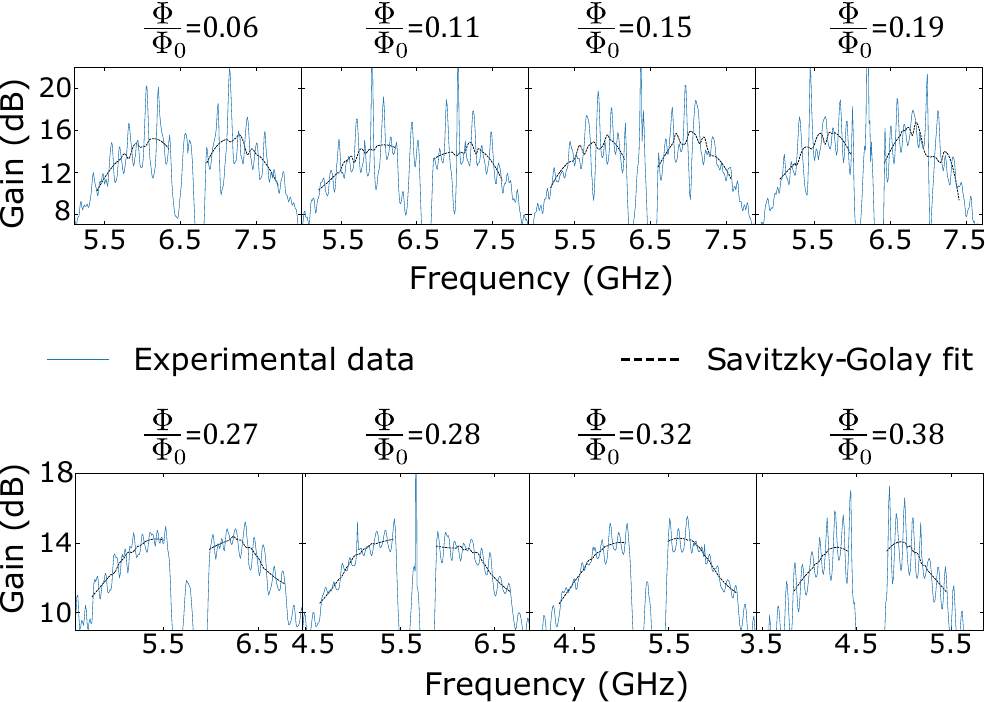}
\caption{Smoothing of several gain profiles using a Savitzky-Golay filter. As the maximum gain stays constant for all the fluxes, there is an improvement in the ripples as the characteristic impedance of the STWPA comes closer to \SI{50}{\ohm}. Note that the y-axis scale has a wider range for the upper row.} 
\label{SavgolGain}
\end{figure}

In this section, we explain how we extracted the root mean square value of the gain ripples for sample B. As explained in Sec.~\ref{sec:flux}, we fitted the different gain profile with a Savitzky-Golay filter in order to smooth it, on a \SI{-3}{\decibel} band as shown in~\cref{SavgolGain} for eight different magnetic fluxes. We then defined the standard deviation as:
\begin{equation}\label{W}
W= \sqrt{\frac{\Sigma_\text{f}(G_\text{exp,f} - G_\text{smooth,f})^{2}}{N}}
\end{equation}
W is plotted as the RMS value of the gain ripples in the main text for more than 25 different gain profiles taken at different magnetic fluxes. The RMS value is plotted as a function of the small-signal characteristic impedance $Z_\text{c}(\Phi)=\sqrt{\frac{\bar{L}(\Phi)}{\bar{C}^\text{g}}}$. $\bar{C}^\text{g}$ is kept the same for all magnetic fluxes while $\bar{L}(\Phi)$ is inferred by fitting the dispersion relation for each corresponding flux. 

%

\end{document}